\documentstyle[psfig]{mn}

\newif\ifAMStwofonts

\ifoldfss
  \ifCUPmtlplainloaded \else
    \NewTextAlphabet{textbfit} {cmbxti10} {}
    \NewTextAlphabet{textbfss} {cmssbx10} {}
    \NewMathAlphabet{mathbfit} {cmbxti10} {} 
    \NewMathAlphabet{mathbfss} {cmssbx10} {} 
  \fi
  \ifAMStwofonts
    \ifCUPmtlplainloaded \else
      \NewSymbolFont{upmath} {eurm10}
      \NewSymbolFont{AMSa} {msam10}
      \NewMathSymbol{\upi}     {0}{upmath}{19}
      \NewMathSymbol{\umu}     {0}{upmath}{16}
      \NewMathSymbol{\upartial}{0}{upmath}{40}
      \NewMathSymbol{\leqslant}{3}{AMSa}{36}
      \NewMathSymbol{\geqslant}{3}{AMSa}{3E}

    \fi
  \fi
\fi 

\ifnfssone
  \newmathalphabet{\mathit}
  \addtoversion{normal}{\mathit}{cmr}{m}{it}
  \addtoversion{bold}{\mathit}{cmr}{bx}{it}
  \newmathalphabet{\mathbfit} 
  \addtoversion{normal}{\mathbfit}{cmr}{bx}{it}
  \addtoversion{bold}{\mathbfit}{cmr}{bx}{it}
  \newmathalphabet{\mathbfss} 
  \addtoversion{normal}{\mathbfss}{cmss}{bx}{n}
  \addtoversion{bold}{\mathbfss}{cmss}{bx}{n}
  \ifAMStwofonts
    \ifCUPmtlplainloaded \else
      \UseAMStwoboldmath
      \makeatletter
      \new@mathgroup\upmath@group
      \define@mathgroup\mv@normal\upmath@group{eur}{m}{n}
      \define@mathgroup\mv@bold\upmath@group{eur}{b}{n}
      \edef\UPM{\hexnumber\upmath@group}
      \new@mathgroup\amsa@group
      \define@mathgroup\mv@normal\amsa@group{msa}{m}{n}
      \define@mathgroup\mv@bold\amsa@group{msa}{m}{n}
      \edef\AMSa{\hexnumber\amsa@group}
      \makeatother
      \mathchardef\upi="0\UPM19
      \mathchardef\umu="0\UPM16
      \mathchardef\upartial="0\UPM40
      \mathchardef\leqslant="3\AMSa36
      \mathchardef\geqslant="3\AMSa3E
    \fi
  \fi
\fi 

\ifnfsstwo
  \DeclareMathAlphabet{\mathbfit}{OT1}{cmr}{bx}{it}
  \SetMathAlphabet\mathbfit{bold}{OT1}{cmr}{bx}{it}
  \DeclareMathAlphabet{\mathbfss}{OT1}{cmss}{bx}{n}
  \SetMathAlphabet\mathbfss{bold}{OT1}{cmss}{bx}{n}
  \ifAMStwofonts
    \ifCUPmtlplainloaded \else
      \DeclareSymbolFont{UPM}{U}{eur}{m}{n}
      \SetSymbolFont{UPM}{bold}{U}{eur}{b}{n}
      \DeclareSymbolFont{AMSa}{U}{msa}{m}{n}
      \DeclareMathSymbol{\upi}{0}{UPM}{"19}
      \DeclareMathSymbol{\umu}{0}{UPM}{"16}
      \DeclareMathSymbol{\upartial}{0}{UPM}{"40}
      \DeclareMathSymbol{\leqslant}{3}{AMSa}{"36}
      \DeclareMathSymbol{\geqslant}{3}{AMSa}{"3E}
    \fi
  \fi
\fi 

\ifCUPmtlplainloaded \else
  \ifAMStwofonts \else 
    \def\upi{\pi}
    \def\umu{\mu}
    \def\upartial{\partial}
  \fi
\fi

\title{Detection of Weak Gravitational Lensing by Large-scale Structure}
\author[D. J. Bacon et al.]
{David~J.~Bacon,$^1$\thanks{E-mail: djb@ast.cam.ac.uk}
Alexandre~R.~Refregier$^1$ \& Richard~S.~Ellis$^{1,2}$ \\
$^1$ Institute of Astronomy, Madingley Road, Cambridge CB3 OHA, UK \\
$^2$ California Institute of Technology, Pasadena CA 91125, USA\\}

\date{Accepted ---. Received ---; in original form ---.}

\pagerange{\pageref{firstpage}--\pageref{lastpage}}
\pubyear{2000}

\begin{document}

\maketitle

\label{firstpage}

\begin{abstract}
We report a detection of the coherent distortion of faint galaxies
arising from gravitational lensing by foreground structures. This
``cosmic shear'' is potentially the most direct measure of the mass
power spectrum, as it is unaffected by poorly-justified assumptions
made concerning the biasing of the distribution. Our detection is
based on an initial imaging study of 14 separated $8' \times 16'$
fields observed in good, homogeneous conditions with the prime focus
EEV CCD camera of the 4.2m William Herschel Telescope. We detect an
rms shear of 1.6\% in $8'\times8'$ cells, with a significance of
$3.4\sigma$. We carefully justify this detection by quantifying
various systematic effects and carrying out extensive simulations of
the recovery of the shear signal from artificial images defined
according to measured instrument characteristics. We also verify our
detection by computing the cross-correlation between the shear in
adjacent cells. Including (gaussian) cosmic variance, we measure the
shear variance to be $(0.016)^2 \pm (0.012)^2 \pm (0.006)^2$, where
these $1\sigma$ errors correspond to statistical and systematic
uncertainties, respectively. Our measurements are consistent with the
predictions of cluster-normalised CDM models (within $1\sigma$) but a
COBE-normalised SCDM model is ruled out at the $3.0\sigma$ level. For
the currently-favoured $\Lambda$CDM model (with $\Omega_{m}=0.3$), our
measurement provides a normalisation of the mass power spectrum of
$\sigma_{8}=1.5\pm0.5$, fully consistent with that derived from
cluster abundances. Our result demonstrates that ground-based
telescopes can, with adequate care, be used to constrain the mass
power spectrum on various scales. The present results are limited
mainly by cosmic variance, which can be overcome in the near future
with more observations.
\end{abstract}


\section{Introduction}
\label{intro}
Determining the large scale distribution of matter
remains a major goal of modern cosmology. Comparisons between
theory and observations are hampered fundamentally by the fact
that the former is concerned with dark matter whereas the latter
usually probes luminous matter, particularly when the distribution
is probed by galaxies and clusters. By contrast, gravitational
lensing provides direct information concerning the total mass
distribution, independently of its state and nature. As a result,
lensing has had considerable impact in studies of cluster mass
distributions (see reviews by Fort \& Mellier 1994, Schneider
1996) and observational limits have improved significantly. Weak
shear has now been detected $>$1.5 Mpc from the centre of the
cluster Cl0024+16 (Bonnet et al. 1994), and in a supercluster
(Kaiser et al. 1998).

Weak lensing by large-scale structure also produces small coherent
distortions in the images of distant field galaxies (see Mellier
1999; Kaiser 1999; Bartelmann \& Schneider 1999 for recent
reviews). A measurement of this effect on various scales (defined
as `cosmic shear') would provide invaluable cosmological
information (Kaiser 1992; Jain \& Seljak 1997; Kamionkowski et al.
1997; Kaiser 1998; Hu \& Tegmark 1998; Van Waerbeke et al. 1998).
In particular, it would yield a direct measure of the power
spectrum of density fluctuations along the line of sight and thus
provide an independent constraint on large scale structure models
and cosmological parameters.

Because of its small amplitude (a few percent on arcmin scales for
favoured CDM models), cosmic shear has however been difficult to
detect. In a pioneering paper, Mould et al. (1994) attempted to
detect the coherent distortion of $R\sim$26 field galaxies over a
9.6 arcmin diameter field and found an upper limit quoted in terms of the
{\it rms} shear at the 4\% level. A search for this effect is the
object of active observational effort (Van Waerbeke et al. 1998;
Refregier et al. 1998; Seitz et al. 1998; Rhodes et al. 1999;
Kaiser 1999). At present however, no unambiguous detections of
cosmic shear have been reported (see however the limited results of
Villumsen 1995; Schneider et al. 1998).

A fundamental limitation of narrow field imaging as a probe of cosmic
shear is that arising from cosmic variance, i.e. the fluctuation in
the lensing signal measured with a limited number of pencil beam sight
lines. Only through the analysis of image fields in many
statistically-independent directions can this variance be
overcome. Prior to such a measurement, it is important to demonstrate
a reliable detection strategy, particularly in the presence of
significant instrumental and other systematic effects.

In this paper, we report the detection of a cosmic shear signal
with 14 separated $16' \times 8'$ fields observed with the 4.2m
William Herschel Telescope (WHT). We provide a detailed treatment
of systematic effects and of the shear measurement method. We test
our results with numerical simulations of lensed images and
quantify both our statistical and systematic errors. We discuss
the consequence of our measurement for the normalisation of the
mass power spectrum. Subsequent papers will extend this technique
to a larger number of fields, reducing the limitations caused by
cosmic variance.

This paper is organised as follows. In \S\ref{theory}, we introduce
the theory of weak lensing in the context of a cosmic shear survey. In
\S\ref{survey} we discuss our observational strategy for detecting it
and describe our observations taken at the WHT and the routine aspects
of data reduction. In \S\ref{object}, we describe the generation of
the object catalogue and how the image parameters were measured. In
\S\ref{distortion} we discuss and characterise distortions introduced
by the telescope optics. In \S\ref{shear_method} we discuss the point
spread function and present our shear measurement method, alongside an
important comparison with the same analysis conducted with simulated
data (\S\ref{simulation}). In \S\ref{estimator}, we describe the
estimator used for measuring the shear variance and the
cross-correlation between adjacent cells. In \S\ref{results}, we
present our results. Our conclusions are summarised in
\S\ref{conclusion}.

\section[]{Theory}
\label{theory}
\subsection{Distortion Matrix}
Gravitational lensing by large scale structure produces
distortions in the image of background galaxies (see Mellier 1999;
Kaiser 1999; Bartelmann \& Schneider 1999 for recent reviews).
These distortions are weak (about 1\%) and can be fully
characterised by the distortion matrix
\begin{equation}
\label{eq:psi_def_theory}
\Psi_{ij} \equiv \frac{\partial (\delta \theta_{i})}{\partial \theta_{j}}
  \equiv \left( \begin{array}{cc}
\kappa +\gamma_{1} & \gamma_{2}\\
\gamma_{2} & \kappa - \gamma_{1} \\
\end{array} \right),
\end{equation}
where $\delta \theta_{i}({\mathbf \theta})$ is the displacement
vector produced by lensing on the sky. The convergence $\kappa$
describes overall dilations and contractions. The shear
$\gamma_{1}$ ($\gamma_{2}$) describes stretches and compressions
along (at $45^{\circ}$ from) the x-axis.

The distortion matrix is directly related to the matter density
fluctuations along the line of sight by
\begin{equation}
\label{eq:psi_dchi}
\Psi_{ij} = \int_{0}^{\chi_{h}} d\chi ~g(\chi) \partial_{i}
   \partial_{j} \Phi
\end{equation}
where $\Phi$ is the Newtonian potential, $\chi$ is the comoving
distance, $\chi_{h}$ is the comoving distance to the horizon, and
$\partial_{i}$ is the comoving derivative perpendicular to the line
of sight. The radial weight function $g(\chi)$ is given by
\begin{equation}
g(\chi) = 2 \int_{\chi}^{\chi_{h}} d\chi'~n(\chi')
   \frac{r(\chi)r(\chi'-\chi)}{r(\chi')},
\end{equation}
where $r$ is the comoving angular diameter distance, and $n(\chi)$ is
the probability of finding a galaxy at comoving distance $\chi$ and is
normalised as $\int d\chi n(\chi) =1$. If the galaxies all lie at a
single distance $\chi_{s}$, $n(\chi)=\delta(\chi-\chi_{s})$ and
\begin{equation}
g(\chi) = 2 \frac{r(\chi) r(\chi_{s}-\chi)}{r(\chi_{s})}
\end{equation}

In practice, this approximation is accurate to within 10\%, if
$\chi_{s}$ is set to the median distance of the galaxy sample.
This is adequate given the median redshift of our galaxy sample is
itself uncertain by about 25\% (see \S\ref{zref}), yielding an
uncertainty in the predicted rms shear of about 20\% (see
Eq.~[\ref{eq:sigg_sig8}] below).

\subsection{Power Spectrum}
The amplitude of the cosmic shear can be quantified statistically by
computing its 2-dimensional power spectrum (Jain \& Seljak 1997;
Kamionkowski et al. 1997; Schneider et al. 1997; Kaiser 1998). For
this purpose, we consider the Fourier transform of the shear field
\begin{equation}
\widetilde{\gamma_{i}}({\mathbf l}) = \int d^{2}\theta
  ~\gamma_{i}({\mathbf \theta}) e^{i {\mathbf l \cdot \theta}}
\end{equation}
The shear power spectrum $C_{{\mathbf l}}^{ij}$ is defined by
\begin{equation}
\label{eq:clij_def}
\langle \widetilde{\gamma_{i}}({\mathbf l})
        \widetilde{\gamma_{j}}({\mathbf l'}) \rangle
= (2\pi)^{2} \delta^{(2)}({\mathbf l} - {\mathbf l'}) C_{{\mathbf l}}^{ij}
\end{equation}
where $\delta^{(2)}$ is the 2-dimensional Dirac-delta function, and
the brackets denote an ensemble average.  It is also useful to define
the scalar power spectrum $C_{l} = C_{{\mathbf l}}^{11} + C_{{\mathbf
l}}^{22}$ for the shear amplitude by
\begin{equation}
\langle \widetilde{\gamma_{i}}({\mathbf l})
        \widetilde{\gamma_{i}}({\mathbf l'}) \rangle
= (2\pi)^{2} \delta^{(2)}({\mathbf l} - {\mathbf l'}) C_{{\mathbf l}},
\end{equation}
where the summation convention was used.

Applying Limber's equation in Fourier space (Kaiser 1998) to
Equation~(\ref{eq:psi_dchi}) and using the Poisson equation, we can
express the shear power spectrum $C_{l}$ in terms of the 3-dimensional power
spectrum $P(k,\chi)$ of the mass fluctuations $\delta \rho/\rho$ and
obtain
\begin{equation}
C_{l} = \frac{9}{16} \left( \frac{H_{0}}{c} \right)^{4} \Omega_{m}^{2}
  \int_{0}^{\chi_h} d\chi~\left[ \frac{g(\chi)}{a r(\chi)} \right]^{2}
  P\left(\frac{l}{r}, \chi\right),
\end{equation}
where $a$ is the expansion parameter, and $H_{0}$ and $\Omega_{m}$
are the present value of the Hubble constant and matter density
parameter, respectively. After noting that $C_{l}$ is also equal
to the power spectrum of the convergence $\kappa$, we find that
this expression agrees with that of Schneider et al. (1997). The
component-wise power spectrum is given by
\begin{equation}
C_{{\mathbf l}}^{ij} = u_{i}(\lambda) u_{j}(\lambda) C_{l}
\end{equation}
where $u_{i}(\lambda)=\{\cos(2\lambda),\sin(2\lambda)\}$ and $\lambda$
is the angle of the vector ${\mathbf l}$, counter-clockwise from the
$l_{1}$-axis.

\begin{figure}
\psfig{figure=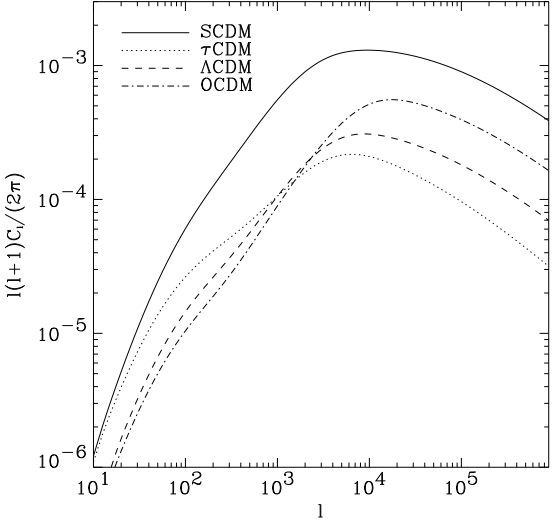,width=80mm} \caption{Shear power spectrum for
each cosmological model and for sources at $z_{s}=1$.  Note that
the SCDM spectrum is larger due to its higher normalisation.}
\label{fig:cl}
\end{figure}

A measurement of the power spectrum enables differentiation
between the different cosmological models listed in
Table~\ref{tab:models}. Standard Cold Dark Matter (SCDM) is
approximately COBE-normalised (Bunn \& White 1997), while the
other variants are approximately cluster normalised ($\sigma_{8}
\Omega_{m}^{.53} =0.6\pm0.1$; Viana \& Liddle 1996). For each
model we compute the non-linear power spectrum using the fitting
formula of Peacock \& Dodds (1996). The resulting power spectra
are shown in Figure~\ref{fig:cl} for sources observed at
$z_{s}=1$.

\begin{table*}
 \centering
 \begin{minipage}{140mm}
  \caption{Cell-averaged statistics for each cosmological model 
             (with $z_{s}=1$)}
  \label{tab:models}
  \begin{tabular}{crrrrrrrr}
\hline
   Model     &  $\Omega_{m}$ & $\Omega_{\Lambda}$ & $\sigma_{8}$ &
 $\Gamma$ & $\sigma_{\gamma}$ (\%) & $\sigma_{\times}$ (\%) &
 $\sigma_{\times 1}$ (\%) & $\sigma_{\times 2}$ (\%) \\
\hline
SCDM & 1.0 & 0 & 1 & 0.50 & 2.60 & 1.62 & 1.23 & 1.05 \\
$\tau$CDM & 1.0 & 0 & 0.6 & 0.25 & 1.25 & 0.86 & 0.64 & 0.58 \\
$\Lambda$CDM & 0.3 & 0.7 & 1 & 0.25 & 1.15 & 0.71 & 0.54 & 0.46\\
OCDM & 0.3 & 0 & 1 & 0.25 & 1.04 & 0.62 & 0.48 & 0.39\\
\hline
\end{tabular}
\end{minipage}
\end{table*}

\subsection{Cell-Averaged Statistics}
For our measurement, we will consider statistics of the shear
averaged over angular cells on the sky. This has the advantage of
diminishing the impact of systematic effects (Rhodes et al. 1999)
and allows extension in later surveys to minimise cosmic variance.
The average shear $\overline{\gamma}_{i}$ in a cell can be written
as
\begin{equation}
\overline{\gamma}_{i} = \int d^{2}\theta~ W({\mathbf
  \theta}) \gamma_{i}({\mathbf \theta})
\end{equation}
where $W({\mathbf \theta})$ is the cell window function and is
normalised as $\int d^{2}\theta~ W({\mathbf \theta}) = 1$.

It is convenient to define the Fourier transform of the window
function as
\begin{equation}
\widetilde{W}_{{\mathbf l}} = \int d^{2}\theta~ W({\mathbf \theta})
 e^{i {\mathbf l \cdot \theta}}.
\end{equation}
For a square cell of side $\alpha$, this is
\begin{equation}
\widetilde{W}_{{\mathbf l}} =
\left( \frac{\sin(\alpha l_{1})}{\alpha l_{1}} \right)
\left( \frac{\sin(\alpha l_{2})}{\alpha l_{2}} \right),
\end{equation}
To a good approximation, we can ignore the small azimuthal dependence
of the window function and approximate
\begin{equation}
\widetilde{W}_{l} \simeq \left( \frac{\sin( \alpha l /
\sqrt{2})} { \alpha l/ \sqrt{2}} \right)^{2}.
\end{equation}

Let us consider 2 cells separated by an angle ${\mathbf \theta}$.
We are interested in the correlation function
\begin{equation}
w_{ij}({\mathbf \theta}) \equiv \langle \overline{\gamma}_{i}(0)
 \overline{\gamma}_{j}({\mathbf \theta}) \rangle
\end{equation}
As is the case in our experiment, we take the separation vector
${\mathbf \theta}$ to lie along the $\theta_{1}$-axis (or equivalently
along the $\theta_{2}$-axis). By taking Fourier transforms and
using Equation~(\ref{eq:clij_def}), we thus obtain
\begin{eqnarray}
w_{ij}({\mathbf \theta}) & \simeq &  \frac{1}{4\pi} \int_{0}^{\infty}
  dl~ l C_{l} \left| \widetilde{W}_{l} \right|^{2} \times \nonumber \\
 & &  \left( \begin{array}{cc}
J_{0}(l \theta)+J_{4}(l \theta) & 0 \\
0 & J_{0}(l \theta)-J_{4}(l \theta) \\ \end{array} \right).
\end{eqnarray}
As noted above, we have ignored the azimuthal dependence of the window
function $\widetilde{W}_{l}$. In particular, the shear variance
$\sigma_{\gamma}^{2} \equiv \langle \overline{\gamma}^{2} \rangle =
w_{11}(0)+w_{22}(0)$ is given by
\begin{equation}
\sigma_{\gamma}^{2} = \frac{1}{2\pi} \int_{0}^{\infty}
  dl~ l C_{l} \left| \widetilde{W}_{l} \right|^{2}.
\end{equation}
We will denote the component-wise covariances between two adjacent
cells by \begin{equation}
\label{eq:sigma_cross}
\sigma_{\times 1}^{2} \equiv w_{11}(\alpha), ~~~\sigma_{\times 1}^{2}
\equiv w_{22}(\alpha),
\end{equation}
and their modulus by $\sigma_{\times}^{2} \equiv \sigma_{\times
1}^{2}+ \sigma_{\times 2}^{2}$. The values of these statistics for
each model are listed in Table~\ref{tab:models} for our cell size
of $\alpha=8'$. The rms shear is of the order of 1\% for the
cluster-normalised models and of about 2\% for the COBE-normalised
model. The cross-correlation rms is about half the zero-lag value
(c.f. Schneider et al. 1997).

\begin{figure}
\psfig{figure=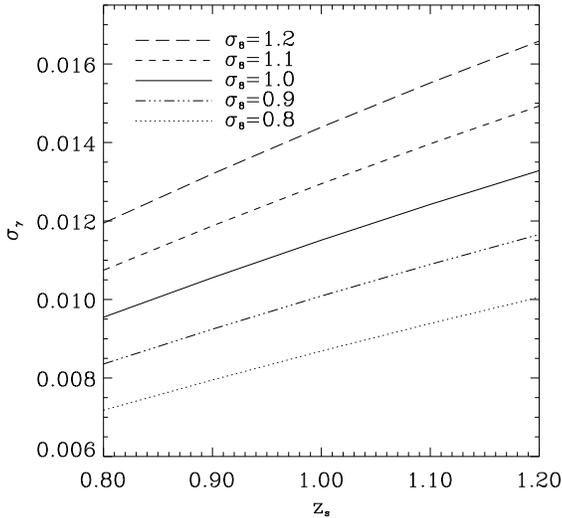,width=80mm}
\caption{Dependence of the rms shear on the source redshift $z_{s}$
and the power spectrum normalisation $\sigma_{8}$. The cell was
chosen to be a square of side $\alpha=8'$.}
\label{fig:sigmag}
\end{figure}

Figure~\ref{fig:sigmag} shows the dependence of $\sigma_{\gamma}$
on the source redshift $z_{s}$ and $\sigma_{8}$ for the
$\Lambda$CDM model (again for $\alpha = 8'$). The range chosen
approximately reflects the likely uncertainty in these parameters
for our experiment. Importantly, the rms shear is more sensitive
to $\sigma_{8}$. A 10\% uncertainty in the source redshift results
in a 8\% uncertainty in $\sigma_{\gamma}$. For this model, the
dependence of $\sigma_{\gamma}$ is very well approximated by
\begin{equation}
\label{eq:sigg_sig8}
\sigma_{\gamma} \simeq  0.0115 z_{s}^{0.81} \sigma_{8}^{1.25},
\end{equation}
in agreement with the scaling laws of Jain \& Seljak (1997).

\section[]{Data}
\label{survey}
\subsection{Survey Strategy}

In order to detect and ultimately measure the cosmic shear, an
array of deep imaging fields is required. These must be randomly
placed on the sky to provide a fair sample, and 
should be well separated in order to be statistically independent,
from the point of view of cosmic variance. As mentioned in
\S\ref{intro}, it is expedient to distinguish between a {\em
detection} based on a careful analysis of a few fields, noting
carefully the systematic effects, before embarking upon an
exhaustive {\em measurement} survey utilising a larger number of
fields to beat down the uncertainties arising from cosmic
variance. With these factors in mind, we now discuss our strategy
and observations using the William Herschel Telescope (WHT).

A bank of appropriate fields were selected for observation with the
WHT prime focus CCD Camera (field of view 8' $\times$ 16', pixel size
0.237'', EEV CCD) in the $R$ band. This photometric band offers the
deepest imaging for a given exposure time with minimal
fringing. Fields were selected using the Digital Sky Survey by
choosing coordinates randomly within the range appropriate for the
time of observations.  Each field was retrospectively checked to see
whether it contained large galaxies ($\gg 5$ arcsec)
(which would occult a significant fraction of the imaging field) or
prominent groups/clusters (located using the NASA/IPAC Extragalactic
Database) on a scale comparable to that under study ($\simeq$8
arcmin). There is, of course, a danger of over-compensating by
exclusion in this respect but, fortunately, none of the
originally-chosen fields were discarded according to the above
criteria.

The fields were further required to be $>5^\circ$ away from one other,
in order to ensure statistical independence (c.f. figure
\ref{fig:cl}, where the power is small for $l < 10$).

Using the APM and GCC catalogues, we ensured that the fields
contained no stars with $R<11$ (in order to avoid large areas of
saturation and ghost images). On the other hand, we required the
fields to contain $\simeq 200$ stars with $R<22$ in order to map
carefully the anisotropic PSF and the camera distortion across the
field of view. In order to achieve this, the fields were chosen to
be at intermediate Galactic latitudes ($30^\circ < b < 70^\circ$; see
Table \ref{tab:coords}). A calibration of the stellar density
at limits fainter than the APM and GCC catalogues was obtained
from a test WHT image (see below).

The final constraint on field position was our desire to observe
each field within $20^\circ$ of the telescope's zenith during the
observing run; this reduces image distortion introduced by
telescope and instrument flexure. This criterion was relaxed for
the fields VLT1, CIRSI1 and CIRSI2 (see nomenclature below).

Table \ref{tab:coords} summarises the positions and Galactic latitude
of the fields which are used in this paper. Two fields are in common
with the VLT (Mellier et al., in preparation) and HST STIS (Seitz et
al. 1998) cosmic shear programmes, allowing future comparisons with
these programmes.  A further two fields spanned the Groth Strip (Groth
et al. 1998; Rhodes 1999) a deep survey conducted with HST, which has
previously been studied for cosmic shear detection (Rhodes et al.
1999). Finally, two fields were chosen to be in common with the current
CIRSI photometric redshift survey (Firth et al., in preparation) to
give us clearer understanding of the redshift distribution of objects
in our fields at a later date.

\begin{table*}
 \centering \begin{minipage}{170mm}
\caption{Field Coordinates (equinox 2000) and Properties}
\label{tab:coords}
 \begin{tabular}{crrrrrrr} \hline
Field name & RA (h:m:s) & Dec (d:m:s) & Galactic & Seeing   &
Magnitude & Median & No. Survey Galaxies \\
           &            &             & latitude & (arcsec)
& limit & magnitude & $(8' \times 16' field$)\\
           &            &             & (deg)    &        & (imcat $5\sigma)$
&  of survey     \\ & & & & & & galaxies\\
\hline
WHT0 & 02:03:09.31 & 11:30:20.0 & -47.6 & 0.59 & 26.2 & 23.1 & 1550\\
WHT3   & 14:00:15.00 &  10:13:40.0 &  66.6 & 0.82 & 26.2 & 23.3 & 2141\\
WHT5   & 14:50:46.67 &  20:18:03.2 &  61.9 & 0.76 & 26.5 & 23.7 & 2181\\
WHT7   & 15:13:40.86 &  36:31:30.8 &  58.6 & 0.83 & 25.9 & 23.0 & 1354\\
WHT11  & 16:31:44.28 &  27:56:30.0 & 41.6 & 0.85 & 26.0 & 23.3 & 1379\\
WHT12  & 16:37:20.00 & 20:46:30.0 & 38.4 & 0.90 & 26.0 & 23.3 & 1855\\
WHT14  & 16:51:15.38 & 25:46:44.0 &  36.8 & 0.99 & 25.9 & 23.2 & 1701\\
WHT16  & 17:13:40.00 &  38:39:19.0 &  34.9 & 0.78 & 25.8 & 23.4 & 2074\\
WHT17  & 14:24:38.10 &  22:54:01.0 &  68.5 & 0.63 & 27.3 & 24.5 & 2287\\
VLT1   & 12:28:18.50 &  02:10:05.0 &  64.4 & 0.71 & 26.4 & 23.6 & 1721\\
VLT2   & 15:28:43.00 &  10:14:20.0 &  49.3 & 0.79 & 26.1 & 23.4 & 2093\\
CIRSI1 & 12:05:35.01 & -07:43:00.0 & 60.1 & 1.14 & 25.4 & 22.6 & 1192\\
CIRSI2 & 15:23:37.00 & 00:15:00.0 & 60.4 & 0.76 & 26.3 & 23.5 & 1824\\
GROTH1 & 14:17:18.74 & 52:20:18.5 &  53.4 & 0.78 & 26.1 & 23.4 & 2237\\
GROTH2 & 14:15:35.00 &  52:08:48.0 &  44.7 & 0.89 & 26.1 & 23.6 & 1195\\
\hline
\end{tabular}
\end{minipage}
\end{table*}


An exposure time of 1 hour on the WHT enables the detection of $R$=26
objects with a signal-to-noise of 5.8 in 0.8'' seeing. This limit
should correspond to a median redshift of about $z_s\simeq$1.2.  In our
eventual analysis, we will introduce a brighter limit so as to keep
only resolved galaxies (referred to as the survey sample). This serves
to reduce the median redshift to about 0.8 (see $\S$9.3).  We
however note from figure \ref{fig:sigmag} that our expected shear
signal is not very sensitive to median redshift ($\sigma_{\gamma}
\propto z_s^{0.8}$). In \S\ref{results}, we will show that the
resulting depth is still sufficient to detect the lensing signal.

\subsection{Observations}
We observed 14 selected fields with the WHT during the nights of 13-16
May 1999. For each field, a total of 4 exposures in $R$, each of 900s,
was taken. All fields were observed as they passed through the
meridian.

Each exposure on a given field was offset by 10'' from its predecessor
in order to remove cosmetic defects and cosmic rays, and to measure
the optical distortion of the telescope and camera (see
\S\ref{distortion} and \S\ref{shear_method}). All but two of the
fields were observed with the long axis of the CCD pointing East-West;
the exception being the two Groth fields, for which a $45^\circ$
rotation (i.e. North-West orientation) was effected to align the WHT
exposures with the HST survey (Groth et al. 1998; Rhodes 1999). Bias
frames and sky flats were taken at the beginning and end of each
night, and standard star observations were interspersed with the
science exposures. The median seeing on our used exposures is 0.81''; 
one exposure with seeing $>1.2$'' was excluded.

Table \ref{tab:coords} lists the seeing and 'imcat' magnitudes
corresponding to $5 \sigma$ detections for each field. An 'imcat'
signal-to-noise (see Section 7) of 5.0 corresponds to a median
$R$=26.1; the median magnitude of galaxies on a field is $R$=25.2.  To
measure the shear, a number of cuts have to be applied to our object
catalog (see \S\ref{shear_method}). Table \ref{tab:coords} also lists
the median magnitude of our final sample. At our final subsample limit
of $R$=23.4, the median redshift is $\simeq 0.8\pm 0.2$
(Cohen et al 2000, $\S$9.3). The median number density of adopted
survey sources is 14.3 arcmin$^{-2}$ (see section \ref{shear_method}).
\label{zref}

In addition to the 13 useful fields, we had already obtained a test
field (WHT0) in service time, and were also kindly given access to a
suitable archival field, WHT17. Both were taken in good conditions:
WHT0 is a 1 hour exposure in the $I$ band, whereas WHT17 is a 1.5 hour
exposure in $R$ (chosen to include a known quasar). Removing these
fields does not significantly alter our results. In terms of uniformity,
apart from the deeper WHT17 field, the standard deviation in limiting
magnitude is $\simeq$0.2 mag which we consider acceptable for our
survey. The error on magnitude zero point derived from standard stars is at
the much lower level of 0.03 mag.


\subsection[]{Data Reduction}
The reduction of these deep images proceeded along a standard
route. A median-combined bias frame was subtracted from the
skyflats and science exposures, and all such exposures were
divided by a median unit-normalised sky-flat. Although the survey
exposures were undertaken in the $R$ band to avoid fringing,
fringing is still detected at a 0.5\% sky level. In order to
remove these fringes, which could potentially introduce structure
into the image ellipticities, all long dithered exposures for a given
night  ($>15$ exposures per night) were stacked without offsetting with a
sigma-clipping  algorithm. This results in a fringe frame mapping the
background fringes but devoid of foreground objects. The
fringe frame for the relevant night was then subtracted
from each science exposure individually, subtracting off the
multiple of the fringe frame found to minimise the rms background
noise. After applying this technique, the fringes are entirely
imperceptible, any residue having an amplitude within the sky
background noise. We experimented with automated and hand- subtraction
of the fringes and verified that this had no noticeable effect on our
shear analysis.

The mean linear astrometric offset (in fractional number of
pixels) between the four exposures was found by producing
SExtractor (Bertin \& Arnoults 1996) catalogues for each exposure,
containing typically 2000-3000 objects. We used the mean offsets
of the matched objects to align the fields. The images were
shifted by the corresponding non-integer number of pixels using
IRAF's imshift routine, taking linear combinations of
neighbouring pixels to effect the non-integer pixel shifts. As
discussed in \S\ref{distortion}, we find no need to rotate the
exposures with respect to each other, or to make further
astrometric distortions to compensate for the optical distortion
of the instrument.

The resulting four exposures for each science field were stacked
with sigma-clipping. Since each exposure is 10'' away from the
others, bad columns and cosmic rays were rejected. The images were
examined visually and remaining defective pixels (e.g. a star just
outside field of view leading to light leakage onto an area of the
CCD; or highly saturated stars) were flagged as potentially
unreliable.

\begin{figure}
\psfig{figure=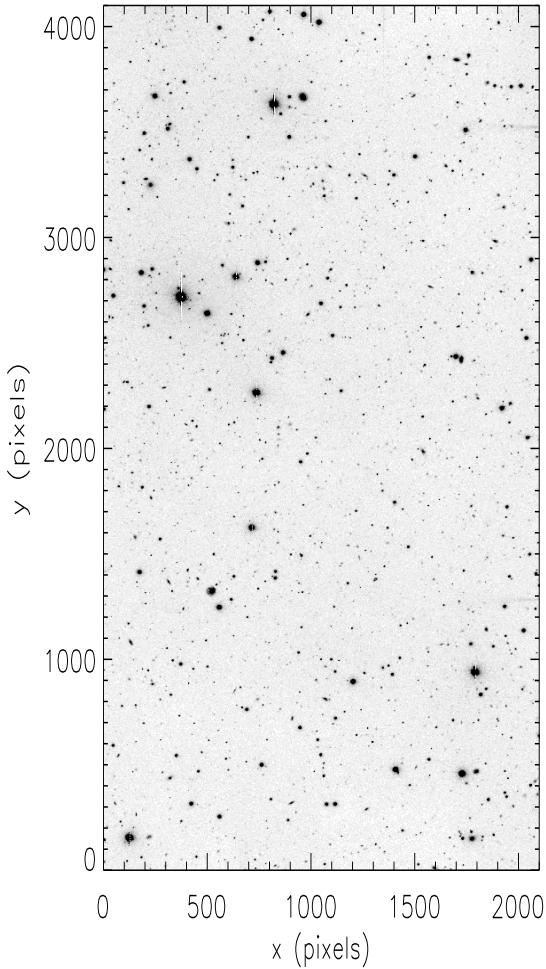,width=80mm,height=140mm} 
\caption{Example
reduced image (CIRSI2); the field of view is 8'$\times$ 16'. Note
that in our analysis, we divide each such field into two 8'
$\times$ 8' cells.} 
\label{fig:cirsi2}
\end{figure}

\section[]{Image Analysis}
\label{object}
We are now ready to measure the ellipticities of the galaxies on
each field, and to apply the necessary corrections in order to
take into account the smearing effect of the atmosphere (`seeing')
plus tracking and other instrumental distortions introduced by the
telescope and camera optics. Only then can we ascertain the true
cosmic shear by averaging the ellipticity distributions of the
corrected galaxies. If no shear were present on a given field, the
mean ellipticity would be zero, within the noise expected from the
non-circularity of galaxies and pixelisation effects. If a shear is
present, the mean ellipticity will be significant, especially when
results are combined from many fields.

A number of methods have recently been proposed to derive the shear
from galaxy shapes (Kaiser et al 1995; Rhodes et al. 1999; Kuijken
1999; Kaiser 1999). Here we choose the most documented method, namely
the KSB formalism proposed by Kaiser et al (1995) and further
developed in Luppino \& Kaiser (1997) and Hoekstra et
al. (1998). While this method is known to have a number of
shortcomings (Rhodes et al. 1999; Kuijken 1999; Kaiser 1999), it is
nevertheless the simplest and is readily available. As we will show in
\S\ref{simulation} using simulations, the method is suitable for our
purposes, after a number of precautions are taken (see Bacon et
al. 2000 for more details). We therefore use this method as provided
by the {\tt imcat} software, a numerical implementation of Kaiser et
al. 1995.

The first task in this process is to detect all objects present on the
fields down to the background noise level, and to measure their
shapes. We then wish to measure their {\em polarisabilities}
i.e. measures of how each is affected by an isotropic smear (due
principally to the atmosphere), an anisotropic smear (due to tracking
errors at the telescope and local coaddition errors due to astrometric
distortion) and shear (both the real gravitational shear and optical
distortions due to the telescope and camera optics). One should note
the distinction between smear and shear: a smear is a convolution of
the image with a kernel, whereas a shear is a stretching of the image
which conserves surface brightness. We will now describe the method for
finding objects, and for measuring their ellipticities and their shear
and smear polarisabilities.

\subsection{Object Detection}

For the purpose of detecting cosmic shear, it is expedient to
divide each of our fields into 2 $8'\times 8'$ cells, since the
signal is stronger on smaller scales (see Figure \ref{fig:cl}).
Furthermore, the mean shear correlation between two adjacent cells
is expected to be about $0.7$\% (see Section 2), and can thus be
used to independently verify our results.

We use the {\tt imcat} software to find objects in each cell, and
to measure their ellipticities, radii, magnitudes, and
polarisabilities. The {\tt hfindpeaks} routine convolves the cell
with Mexican hat functions of varying size, and maximally
significant peaks in surface brightness after convolution are
designated objects. The radius of the hat giving the largest
signal to noise $\nu$ for a given galaxy is attributed to that
galaxy as its filter radius $r_g$. The local sky background is
estimated by the {\tt getsky} routine, and aperture photometry is
carried out on the objects, determining magnitude and half-light
radius $r_h$ for all objects using the {\tt apphot} routine.

\subsection{Shape Measurement}
Using the {\tt getshapes} routine, we then measure the weighted
quadrupole moments of each object which are defined as
\begin{equation}
I_{ij}\equiv \int d^2 x ~ w(x) x_i x_j I(\mbox{\boldmath $x$})
\label{eq:qij}
\end{equation}
where $I$ is the surface brightness of the object, $x$ is angular
distance from object centre, and $w(x)$ is a Gaussian weight function
of scale length equal to $r_g$.  In this fashion we obtain ellipticity
components
\begin{equation}
\label{eq:e_def}
e_{i} \equiv I_{i} / T,
\end{equation}
where
\begin{equation}
I_1 \equiv I_{11}-I_{22}, \;\; I_2 \equiv 2 I_{21}, \;\; T \equiv
I_{11}+I_{22}.
\end{equation}
We can further define $e \equiv (e_1^2+e_2^2)^{\frac{1}{2}}$, where
$e_1 = e \cos{2 \phi}$ and $e_2 = e \sin{2 \phi}$, where $\phi$ is the
position angle associated with the elongation direction of the object
(anticlockwise from x-axis). The trace $T$ of the quadrupole moments
provides a measure for the rms radius $d$ of the object, which we
define as
\begin{equation}
\label{eq:d_def}
d^{2} \equiv \frac{1}{2} (I_{11}+I_{22}) / I_{0},
\end{equation}
where $I_{0} \equiv \int d^2 x ~ w(x) I(\mbox{\boldmath $x$})$
is the flux of the object.

\subsection{Polarisability}

The {\tt imcat} software also enables us to calculate the smear
and shear polarisabilities. In the following, we briefly review
their function. It is possible (see e.g. KSB 95 Appendix) to
calculate the effects of anisotropic smearing, by replacing the
image $I(\mbox{\boldmath $x$})$ in (\ref{eq:qij}) with a convolved
(i.e. anisotropically smeared) image $I^\prime(\mbox{\boldmath
$x$})$ and by finding the effect on the original $e_i$. It is
found that the galaxy ellipticity $e^g_{\rm smeared}$ can be
corrected for the smear as
\begin{equation}
e^g_{\rm corrected} = e^g_{\rm smeared} - P^g_{sm} p,
\label{eq:ecorrect}
\end{equation}
where the ellipticities are understood to denote the relevant
2-component spinor $e_i$, and $p$ is a measure of PSF anisotropy. The
tensor $P^g_{sm}$ is the smear polarisability, a $2\times 2$
matrix with components involving various moments of surface
brightness. Since for stars $e^*_{\rm corrected}=0$, we can set $p =
(P^*_{sm})^{-1} e^*_{\rm smeared}$, and find
\begin{equation}
e^g_{\rm corrected} = e^g_{\rm smeared} - P^g_{sm} (P^*_{sm})^{-1}
e^*_{\rm smeared}
\label{eq:ecorrect2}
\end{equation}
In this fashion, we can correct a galaxy ellipticity for the
effect of anisotropic smearing, using the smear polarisability
$P^g_{sm}$.

In a similar manner, we can calculate the effect of a {\em
shear}, however it is induced.
Replacing the image $I(\mbox{\boldmath $x$})$ in
(\ref{eq:qij}) with a weakly sheared image, we find that
\begin{equation}
e^g_{\rm sheared} = e^g_{\rm initial} + P^g_{sh} \gamma ,
\label{eq:egsheared}
\end{equation}
where $\gamma$ denotes the two component shear
(Eq.~[\ref{eq:psi_def}]), and $P^g_{sh}$ is the shear
polarisability, a $2\times 2$ matrix with components involving
various moments of surface brightness (different from $P^g_{sm}$
above).

In practice, the lensing shear takes effect before the circular
smearing of the PSF. Luppino and Kaiser (1997) showed that the
{\it pre}-smear shear $\gamma$ averaged over a field can be
recovered using
\begin{equation}
\langle {P_{\gamma}}\gamma  \rangle = \langle {e^g_{\rm corrected}} \rangle
\label{eq:eoverpg}
\end{equation}
where
\begin{equation}
P_{\gamma} = P_{sh}^g - \frac{P_{sh}^*}{P_{sm}^*} P_{sm}^g.
\label{eq:pgamma}
\end{equation}
Here, $e^g_{\rm corrected}$ is the galaxy ellipticity corrected
for smear, as in equation (\ref{eq:ecorrect}), and $P_{sh}^*$ and
$P_{sm}^*$ are the shear and smear polarisabilities calculated for
a star interpolated to the position of the galaxy in question. The
interpretation of the division in this equation is a matter of
debate; our adopted procedure will be found in Section 7. With the
smear and shear polarisabilities calculated by {\tt imcat}, we can
therefore find an estimator for the mean shear in a given cell.

In summary, we can derive a catalogue of objects on a cell. For
every object, we determine its centroid, magnitude, half-light and
filter radii, ellipticity components and polarisabilities as
defined above. We can now use these catalogues to understand and
correct for systematic effects, particularly for instrumental
distortion and PSF-induced effects.

\section[]{Instrumental Distortions}
\label{distortion} The instrumental distortion induced by the
optical system of the telescope must be accounted for. If left
uncorrected, this effect can indeed produce both a spurious shear
and a smearing during the coadding process. In the following, we
first present our method to measure the distortion using dithered
astrometric frames. We then apply this method to our WHT fields
and compare our measured distortion field to that predicted by the
WHT Prime Focus manual (Carter \& Bridges 1995). We then show how
the coadding smear can be computed from the astrometric frames. We
finally quantify the impact of these effects on our lensing
measurement.

\subsection{Measurement of the Astrometric Distortion}

The distortion field introduced by the telescope and camera optics
can be measured from the astrometric shifts of objects observed in
several frames offset by known amounts. Let ${\mathbf x}$ be the
true position of an object. Let ${\mathbf x}^{f}$ be its position
observed in frame $f$, without any correction for the camera
distortion. The observed position can be written as
\begin{equation}
{\mathbf x}^{f} = {\mathbf x} + {\mathbf \delta x}({\mathbf
 x}-{\mathbf \overline{x}}^{f})
\end{equation}
where ${\mathbf \delta x}$ is the displacement produced by the
distortion. The vector ${\mathbf \overline{x}}^{f}$ is the position of
the centre of frame $f$, and can be measured as the average position
of all the objects found in the image. We assume that the displacement
field ${\mathbf \delta x}$ is the same for all frames.

The position of this object observed in another frame $f'$ is
${\mathbf x}^{f'} = {\mathbf x} + {\mathbf \delta x}({\mathbf
x}-{\mathbf \overline{x}}^{f'})$. Here, ${\mathbf \overline{x}}^{f'}$
is the centre position of the new frame, which is assumed to be
displaced from frame $f$ only by a translation. (This formalism can be
easily extended to include a rotation of the frames about their
centre, but this effect is negligible in our case).  If
the offset ${\mathbf \overline{x}}^{f} -{\mathbf \overline{x}}^{f'}$
is small compared to the scale on which ${\mathbf \delta x}$ varies,
we can expand this last expression in Taylor series and get
\begin{equation}
\label{eq:delta_x}
{\mathbf x}^{f'} - {\mathbf x}^{f}  \simeq {\mathbf \Psi}
 ({\mathbf \overline{x}}^{f} -{\mathbf \overline{x}}^{f'}),
\end{equation}
where
\begin{equation}
\label{eq:psi_def}
\Psi_{ij} \equiv \frac{\partial (\delta x_{i})}{\partial x_{j}}
\end{equation}
is the distortion matrix at the location of the object as defined in
Equation~(\ref{eq:psi_def_theory}). Following the lensing
conventions, the distortion matrix can be parametrised as
\begin{equation}
\label{eq:psi_params}
{\mathbf \Psi} \equiv
\left( \begin{array}{cc}
\kappa +\gamma_{1} & \gamma_{2} + \rho \\
\gamma_{2} - \rho & \kappa - \gamma_{1} \\
\end{array} \right),
\end{equation}
where $\kappa$ and $\gamma_{i}$ are the spurious convergence and shear
introduced by the geometrical distortion. We have included the
rotation parameter $\rho$, which, unlike the case of gravitational
lensing, does not necessarily vanish.

The 4 free parameters of the distortion matrix can thus be measured
from the position of an object in 3 frames $f, f'$ and $f''$. This can
be done by solving the system of 4 independent equations formed by
equation~(\ref{eq:delta_x}) and its counterpart for $f$ and $f''$. The
system will not be degenerate, if the offsets ${\mathbf
\overline{x}}^{f} -{\mathbf \overline{x}}^{f'}$ and ${\mathbf
\overline{x}}^{f} -{\mathbf \overline{x}}^{f''}$ are not parallel.

\subsection{Distortion Field for the WHT Prime Focus}

\begin{figure}
\psfig{figure=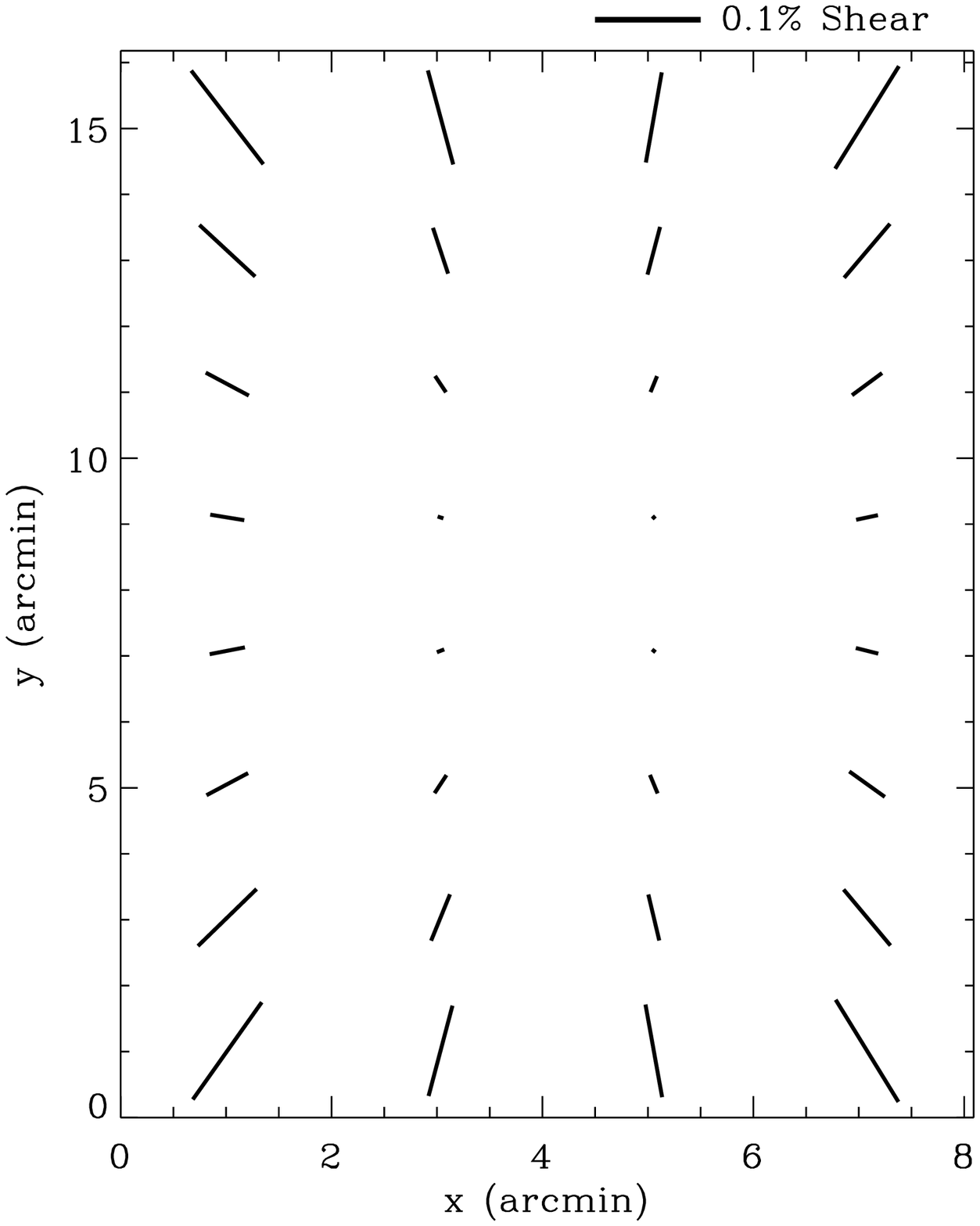,width=80mm}
\psfig{figure=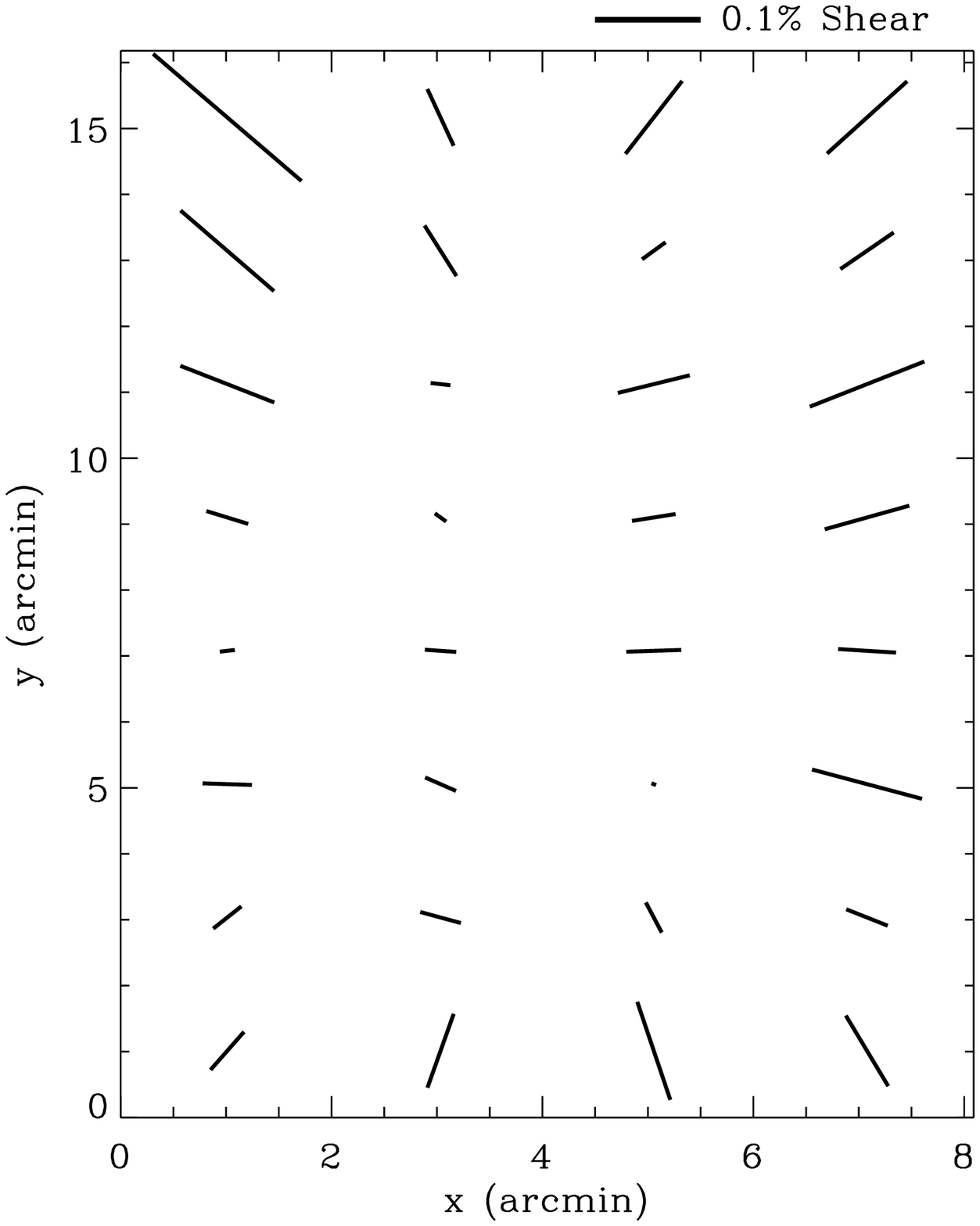,width=80mm}
\caption{Expected (top) and measured (bottom) instrumental shear
pattern for the WHT Prime Focus. The expected pattern was 
derived from the distortion model given in the WHT Prime Focus
manual (Carter \& Bridges 1995). The observed pattern was
measured using 3 astrometric frames in one of our fields.} 
\label{fig:geom}
\end{figure}

First we can compute the expected instrumental distortion using the
specifications in the WHT Prime Focus manual (Carter \& Bridges
1995). The displacement field is expected to be radial with an
amplitude of ${\mathbf \delta x} = a r^{3} {\mathbf \hat{r}}$, where
$r$ is the distance from the optical axis (located at
$(1076.13,2010.7)$ pixels), ${\mathbf \hat{r}} = {\mathbf r}/r$ is the
associated unit radial vector, and $a \simeq 4.27 \times 10^{-10}$
pixels$^{-2}$. Using this expression in
Equations~(\ref{eq:psi_def}-\ref{eq:psi_params}), we can compute the
distortion parameters to be
\begin{eqnarray}\
\label{eq:dist_manual}
\kappa & = & 2 a r^{2} \nonumber \\
\gamma_{i} & = & a r^{2} \hat{e}_{i}^{r} \\
\rho & = & 0, \nonumber
\end{eqnarray}
where $\hat{e}_{i}^{r} \equiv \{r_{1}^{2}-r_{2}^{2},2
r_{1}r_{2}\}/(r_{1}^{2}+r_{2}^{2})$ is the unit radial ellipticity
vector. This therefore predicts a radial instrumental shear with an
amplitude growing like $r^{2}$, reaching $\gamma \sim 0.001$ at the
edge of the chip. This expected shear pattern is shown on figure
\ref{fig:geom}.

Figure~\ref{fig:geom} also shows a typical instrumental shear pattern
measured in one of our fields. This was derived using the method
described above applied to 3 astrometric frames dithered by about 10''
and containing about 15 objects per square arcmin. The uncertainty for
the mean shear component $\gamma_{i}$ in each of the $2'\times2'$ cell
is of about 0.0005. Astrometric measurements thus allow us to measure
the instrumental distortion with very high accuracy.

The measured shear pattern is also approximately radial and agrees
well with the expected pattern. More importantly, it also has an
amplitude of at most 0.001 throughout the field. We have inspected
all of our fields in this manner, and have found only small
field-to-field variations (of about 0.002) for the shear patterns.
In all fields, the maximum instrumental shear is only 0.003 in
single $2'\times2'$ cells.  This number would be even smaller, for
an average over a larger area. We also compared the convergence
$\kappa$ and $\rho$ patterns to that expected from the WHT manual
(Eq~[\ref{eq:dist_manual}]). We again found good agreement with
small field-to-field variations of about 0.002. The origin of
these variations is unknown but could arise perhaps from telescope
flexure. For our purposes, however, it is quite clear that the
instrumental distortion is much smaller than the expected lensing
signal. We therefore neglect this component in the subsequent
analysis.

\subsection{Smear Arising from Co-Addition}

If left uncorrected, instrumental distortions can also produce a
systematic effect on the shapes of galaxies, during the coadding
process. The images of a galaxy from each (distorted) frame will
be slightly offset from one another, and will therefore combine
into a blurred coadded image. Here, we show that this effect is
equivalent to a convolution (or smear) by an additional kernel.
Since this effect will equally affect the stars in the field, it
will be corrected for by the PSF correction described in
\S\ref{shear_method}. It is nevertheless important to estimate the
amplitude of this effect, and to ensure that it does not dominate
the dispersion of the PSF anisotropy.

Let us consider the image of an object which appears on $N_{f}$
frames. As before, let ${\mathbf x}^{f}$ be its centre position on
frame $f$ (after correcting for a translation but not for the
distortion). Let us choose the centre of our coordinate system to
coincide with the centre-of-light ${\mathbf x}^{o} \equiv \sum_{f}
{\mathbf x}^{f} / N_{f}$ of the coadded image. The coadded surface
brightness is then
\begin{equation}
I'({\mathbf x}) = \frac{1}{N_{f}} \sum_{f=1}^{N_{f}} I({\mathbf
x}-{\mathbf x}^{f}),
\end{equation}
where $I({\mathbf x})$ is the (undistorted) surface brightness of the
object, and the factor of $N_{f}^{-1}$ was added for convenience. Note
that the effect of the distortion on the object shape in individual
frames was treated separately in the previous sections, and was thus
ignored in this expression. It is easy to see that $I'$ can be written
as a convolution of $I$ with the kernel
\begin{equation}
Q({\mathbf x}) = \frac{1}{N_{f}} \sum_{f=1}^{N_{f}}
  \delta^{(2)}({\mathbf x}-{\mathbf x}^{f}),
\end{equation}
where $\delta^{(2)}$ is the 2-dimensional Dirac-Delta function.

To estimate the amplitude of the effect, it is convenient and
sufficient to consider the normalised unweighted quadrupole moments
\begin{equation}
J_{ij} \equiv \int d^{2}x ~x_{i} x_{j} I(x) \left/ \int d^{2}x I(x) \right.,
\end{equation}
(see Eq.~[\ref{eq:qij}]) of the undistorted image, and similarly for
the moments $J_{ij}'$ of the coadded image. The unweighted moments of
the kernel $Q({\mathbf x})$ are simply
\begin{equation}
Q_{ij} = \frac{1}{N_{f}} \sum_{f=1}^{N_{f}} x_{i}^{f} x_{j}^{f}
\end{equation}
Because $I$, $I'$ and $Q$ are simply related by a convolution, their
respective quadrupole moments are related by $J'_{ij}=J_{ij}+Q_{ij}$
(see e.g. Rhodes et al. 1999). The rms radius $d'$
(Eq.~[\ref{eq:d_def}]) of the coadded image is thus given by
\begin{equation}
\label{eq:dprime}
d^{\prime 2} = d^{2} + d_{q}^{2},
\end{equation}
where $d$ and $d_{q}$ are the rms radius for the undistorted image and
for the kernel respectively. For simplicity, let us consider an object
which is intrinsically circular. The ellipticity $e_{i}'$ of the coadded
image (see Eq.~[\ref{eq:e_def}]) is then given by ({\it ibid})
\begin{equation}
\label{eq:eprime}
e_{i}'= \frac{d_{q}^{2}}{d^{2}+d_{q}^{2}} e_{i}^{q},
\end{equation}
where $e_{i}^{q}$ is the ellipticity of the kernel.

Turning to the specific case of the WHT observations, let us
consider the ellipticity produced by the coadding smear on a star
observed on 4 frames with a 0.7 arcsec circular seeing. Note that
the effect will be smaller for galaxies which are extended, and so
the following estimate should be considered as an upper limit. For
simplicity, we conservatively assume that the seeing has a
gaussian profile. We inspected all our fields and found that the
astrometric offsets between the different frames was always
smaller than 0.3 pixels. Using Equation~(\ref{eq:dprime}) we
calculated the change $(d'-d)/d$ in the radius of the star which
is always less 2\%, i.e. negligible compared to intrinsic changes
in the seeing size. Using Equation~(\ref{eq:eprime}) we also
computed the induced ellipticity $\epsilon'$ of the star and found
it to be of the order of 0.01 and always less than 0.03. This is
considerably smaller than the rms dispersion in the PSF
ellipticity that we measure in our fields (about 0.07, see
\S\ref{shear_method}) , which must therefore be due to other
effects (tracking errors, atmospheric effects, etc).

Again, we can conclude that smear arising via instrumental
distortions during image coaddition is negligible.

\section{Correction for the Point Spread Function}

\label{shear_method} In order to measure the systematic alignment
of faint background galaxies due to lensing by large-scale structure,
we need to account for more than simply the geometric distortions
discussed in the previous section. We also need to correct for the
effect of the varying atmospheric conditions throughout each exposure
and imperfections in telescope tracking, leading to an anisotropic
smearing of the images. In addition, the isotropic smear arising from
seeing circularises small galaxies, thereby weakening the sought-after
signal. In this section, we first address the anisotropic component of
the contribution to the point spread function (PSF) and then the
isotropic part, thus determining a measure of the true (corrected)
mean shear in each cell.

Although our recipe for measuring the true shear is straightforward,
it is the success of our simulations described in \S\ref{simulation}
which provides justification that our results are reliable at the
necessary 1\% level.

\subsection{Anisotropic Correction}

Our approach is to use equation~(\ref{eq:ecorrect2}) to remove the
anisotropic component of the smearing induced in the galaxy
images. However, we must first remove the extraneous noise detections
in our {\tt imcat} object catalogue, find appropriate well-defined
subcatalogues of stars and galaxies, and generate a functional model
for the stellar ellipticities and polarisabilities over the field of
view.

Firstly, we need to remove noisy detections. We applied a size limit
$r_g > 1.0$ to reject the extraneous very small object detections
which {\tt imcat} finds. We also applied a signal-to-noise $\nu > 15.0$ cut;
see \S\ref{isotropic} for justification of this apparently very
conservative cut. To reduce the noise in our measurement, we also
remove highly elliptical objects with $e > 0.5$.

\begin{figure}
\psfig{figure=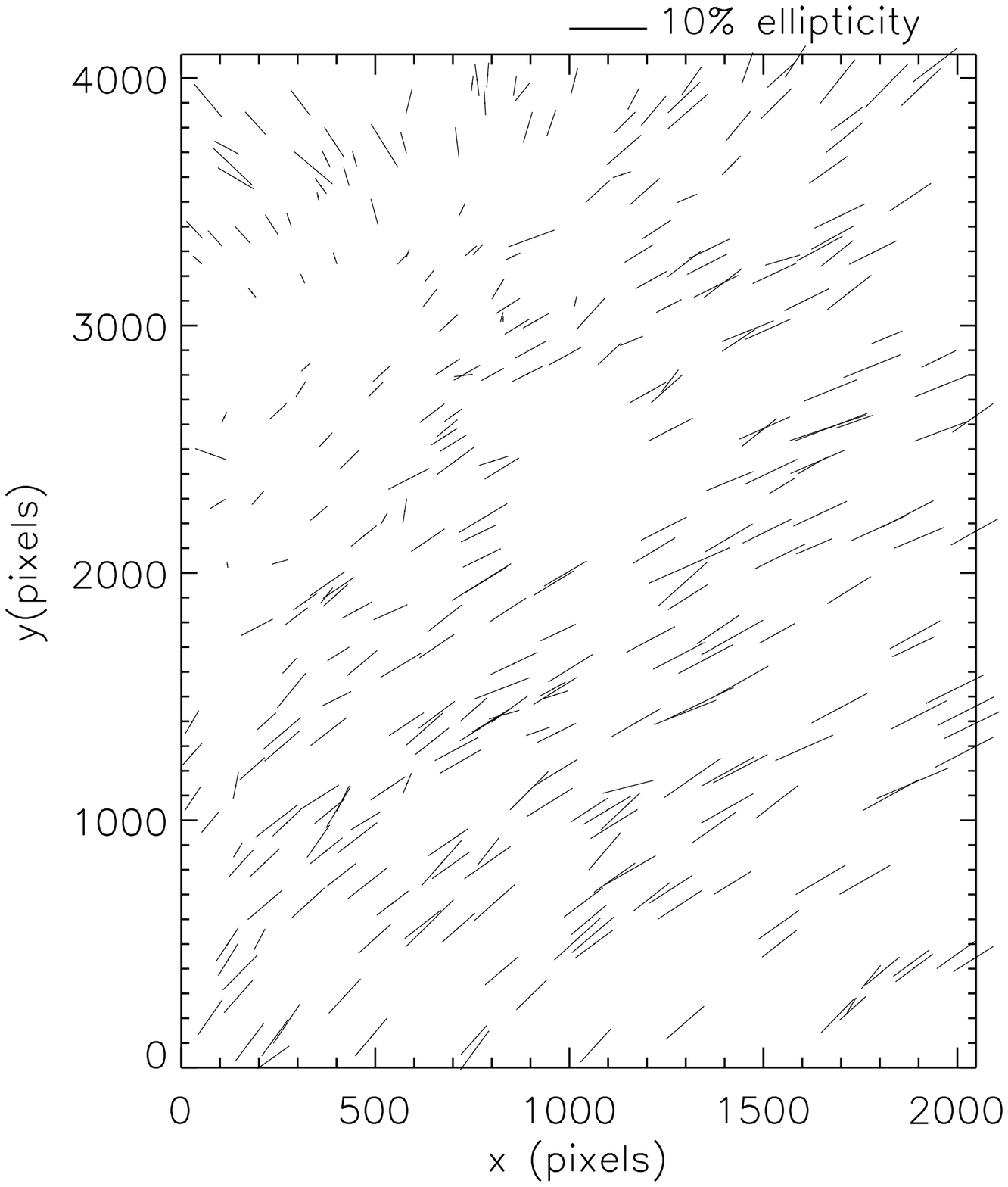,width=80mm,height=110mm}
\psfig{figure=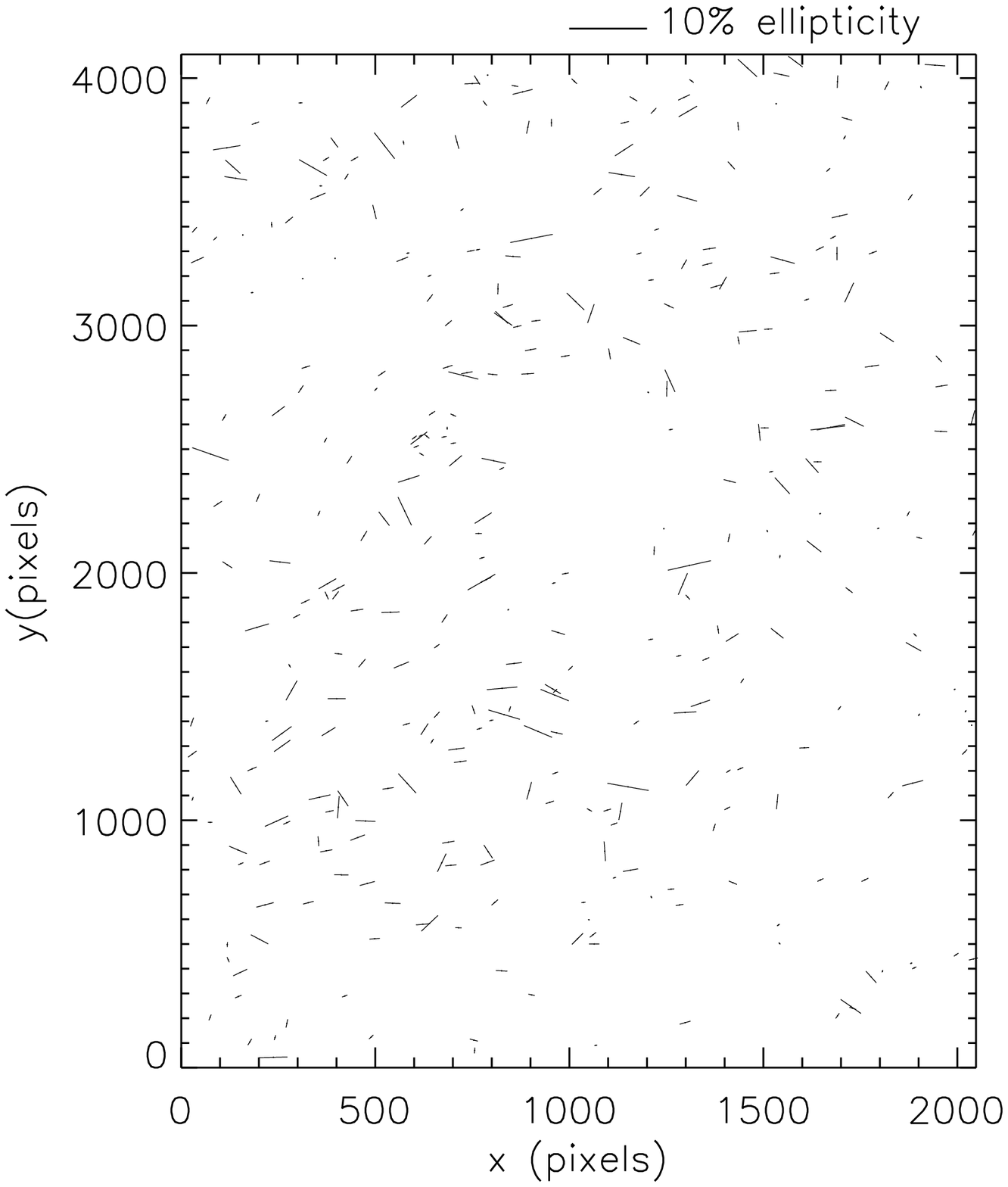,width=80mm,height=110mm}
\caption{Top: Stellar ellipticity distribution for the field of figure
\ref{fig:cirsi2} (CIRSI2). The mean value observed is $\bar{e}^{*} \simeq
0.07$. Bottom: Residual stellar ellipticity after correction. The
residual mean ellipticity is $\bar{e}^{res} \simeq 2.6 \times
10^{-3}$.} \label{fig:starfield}
\end{figure}

Stars were identified using the non-saturated stellar locus on the
magnitude--$r_h$ plane (see figure \ref{fig:magrg}), typically with
$R \simeq$ 19-22. The distribution of stellar ellipticity over a
typical field is shown in figure \ref{fig:starfield}; for this field
we find $e \simeq 0.07$ with only slow positional variations across
the field.  Although the pattern varies from field to field, it is
smooth in all cases. The rms field-to-field stellar ellipticity is
relatively large, $\sigma_{e*} \simeq 0.068$, and must therefore be
removed with care.

In order to use (\ref{eq:ecorrect2}) to correct for these
elongations, we must estimate the positional dependence of stellar
ellipticity and polarisability by interpolation. We adopted an
iterative approach to this problem. We first fitted a 2-D cubic to
the measured stellar ellipticities, plotted the residual
ellipticities $e^{res} = e^* - e^{fit}$ and re-fitted after the
removal of extreme outliers (caused by galaxy contamination,
blended images and noise).

\begin{figure}
\psfig{figure=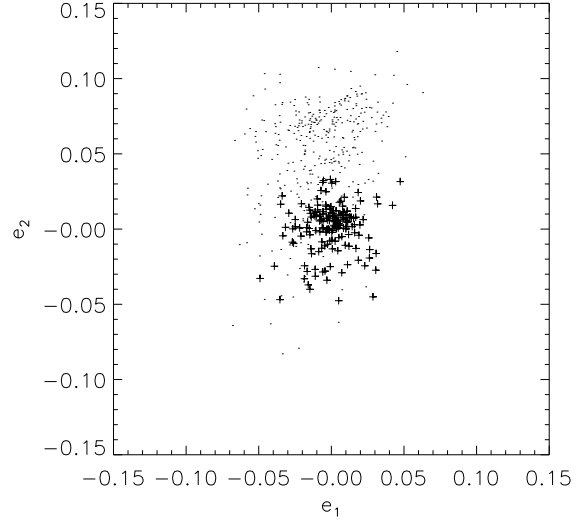,width=80mm} \caption{Effectiveness of
the correction for PSF anisotropy as applied to the field CIRSI2.
Stars initially have a range of ellipticities across the field of
view (dots). After polynomial fitting and correction (see text)
the stars have small mean ellipticity close to zero
($\bar{e}^{res} \simeq 2.6 \times 10^{-3}$; crosses)}
\label{fig:ecorrect}
\end{figure}

Figures \ref{fig:starfield} and \ref{fig:ecorrect} show the stellar ellipticity residual
for the field CIRSI2. Although the mean spurious ellipticity
induced by the instrumental effects is $\bar{e}_1 \simeq -0.009$,
$\bar{e}_2 \simeq 0.052$ over the field, the residual ellipticity
after correction is only $\bar{e}_1^{res}=(0.6\pm 1.2)\times
10^{-3} $, $\bar{e}_2^{res}=(2.5\pm 1.2)\times 10^{-3}$. Despite
the fact that the initial stellar ellipticities on our images are
considerable ($\sigma_{e*} \simeq 0.068$), $e^{res}$ is thus found
to be very small: its field-to-field rms is $\sigma_{e,res} \simeq
1.4 \times 10^{-3}$. This success arises because of the smoothness
of the variation in the stellar ellipticity across each field. The
small residuals will contribute negligibly to the mean shear.

To check the robustness of our anisotropy correction, we used half of
the selected stars on a field, distributed uniformly across the field of
view, to correct the PSF anisotropy; we then compared the final shear 
measurement obtained for this field with that obtained after anisotropy 
correction with the other half of the stars.  We found that the final 
measured shear differed by only 0.1\%.

At this stage we further discard 4 of our cells for which our PSF
interpolation model is not satisfactory. This is due to $r_g^*$ (and
consequently $P_{sm}^*$) showing strong gradients across the cell, or
due to an insufficient number of stars in an area of the cell leading to
poor fitting of the PSF model.

In order to correct galaxies for anisotropic smear, we not
only need the fitted stellar ellipticity field, but also the four
component stellar smear and shear polarisabilities as a function of
position. Here a 2-D cubic is fit for each component of $P_{sm}^*$
and $P_{sh}^*$. Galaxies are then chosen from the magnitude-$r_h$
diagram by removing the stellar locus and objects with $\nu<15$,
$r_{g}<1$, $e>0.5$, as described above. From our fitted stellar
models, we then calculate $e^*$, $P_{sm}^*$ and $P_{sh}^*$ at each
galaxy position, and correct the galaxies for the anisotropic PSF
using equation (\ref{eq:ecorrect2}). As a result, we obtain
$e^{g}_{{\rm corrected}}$ for all selected galaxies in each cell.

\subsection{Isotropic Correction and Shear Measurement}
\label{isotropic}

The final correction arises because of atmospheric seeing which
induces an {\em isotropic} smear. Clearly small objects suffer more
circularisation by the isotropic component of the smear than larger
objects. The goal now is to correct for this bias as well as to
convert from corrected ellipticities to a measure of the corresponding
shear, using $P_{\gamma}$ as introduced in section 5, in equations
(\ref{eq:egsheared}) to (\ref{eq:pgamma}).

We first calculate $P_{\gamma}$ for the galaxies. We opt to treat
$P_{sh}^*$ and $P_{sm}^*$ as scalars equal to half the trace of
the respective matrices. This is allowable, since the non-diagonal
elements are small and the diagonal elements are equal within the
measurement noise (typical $P_{sm,11,22}^*$ = 0.10,
$P_{sm,12,21}^* < 5\times 10^{-4}$, $P_{sh,00,11}^*$ = 1.1,
$P_{sh,12,21}^* < 0.01$). 

With this simplification, we calculate $P_{\gamma}$ according to
equation (\ref{eq:pgamma}). $P_{\gamma}$ is typically a noisy
quantity, so we fit it as a function of $r_g$. We choose to treat
$P_{\gamma}$ as a scalar, since the information it carries is
primarily a correction for the size of a given galaxy, regardless
of its ellipticity or orientation. We thus plot $P_{\gamma}^{11}$
and $P_{\gamma}^{22}$ together against $r_g$, and fit a cubic to
the combined points. Moreover, since $P_{\gamma}$ is unreliable
for objects with $r_g$ measured to be less that $r_g^*$, we remove
all such objects from our prospective galaxy catalogue. Finally,
we calculate a shear measure for each galaxy as (c.f. Eq.
[\ref{eq:eoverpg}])
\begin{equation}
\gamma^{g} = \frac{e^g}{P_{\gamma}},
\end{equation}
where the $P_{\gamma}$ is the fitted value for the galaxy in question.

Because of pixel noise, a few galaxies yield extreme, unphysical
shears $\gamma^{g}$. To prevent these from unnecessarily dominating
the analysis, we have removed galaxies with $\gamma^{g}>2$.  We then
calculate the mean $\bar{\gamma}=\langle \gamma^{g} \rangle$ and error
in the mean $\sigma[\bar{\gamma}]=\sigma[\gamma^{g}]/\sqrt{N_{g}}$ for
this distribution, giving us an estimate for the mean shear in each
cell and its uncertainty.

\label{anticorr}

\begin{figure}
\psfig{figure=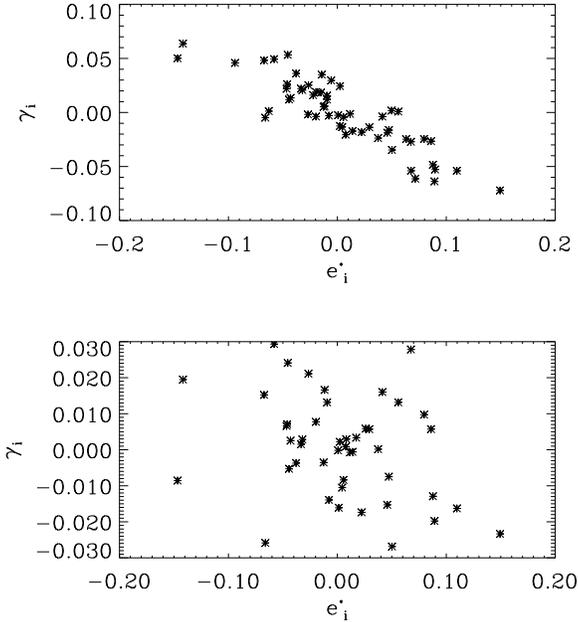,width=80mm} \caption{The anticorrelation of
$\bar{e}_i^*$ and $\bar{\gamma}_i$ plotted for all cells, where
i=1,2 have been superposed, for (top) a $\nu >5$ cut, and (bottom)
a $\nu >15$ cut. Note the trend for $\nu >5$.} \label{fig:anti}
\end{figure}

At this point we encountered an interesting trend. We found that
a signal/noise cut at $\nu > 5$ (as opposed to our conservative $\nu
> 15$) reveals a strong anti-correlation between the mean
shear $\bar{\gamma}_i$ and the mean stellar ellipticity
$\bar{e}^{*}_i$. This can be seen clearly in figure
\ref{fig:anti}. To assess the significance of this effect, we use
the correlation coefficient
\begin{equation}
C_i = \frac {\langle e_i^* \gamma_i \rangle - \langle e_i^* \rangle \langle
\gamma_i \rangle }{\sigma(e_i^*) \sigma(\gamma_i)}.
\end{equation}
For a $\nu > 5$ cut we find $C_1=-0.83$, $C_2=-0.80$, which, for 32
cells, corresponds to a $\gg 3\sigma$ effect. This is clearly
significant, and is due to an over-correction of the PSF for small
galaxies (in Eq.~[\ref{eq:ecorrect2}]). However, for a cut at $\nu >
15$, this reduces to $C_1=-0.31$, $C_2=-0.38$, corresponding to a $1.7
- 2.2 \sigma$ significance for the correlation, which is no longer 
significant. We will take this anticorrelation into account in our
final results.

\section[]{Simulations}
\label{simulation}

In order to verify our analysis, we have conducted a detailed study of
simulated data. The principal aim is to check that the shear we impose
on simulated images is recovered by the detection method described
above in the context of the actual observations. A detailed
description of our simulations will be found in a second paper (Bacon
et al. 2000). Here we describe the relevant details.

We have attempted to create a realistic simulation of a WHT field,
with appropriate counts, magnitudes, ellipticities and diameters
for stars and galaxies, including the effects of seeing, tracking
errors, pixelisation, and an input shear signal.

One approach to this problem would be to directly use sheared Hubble
Space Telescope (HST) images degraded to ground-based
resolution. However, a test of our signal to the required precision
requires an area which is too large to be available in current HST
surveys. We have thus instead chosen a monte-carlo approach, in which
large realisations of artificial galaxy images are drawn to reproduce
the statistics of existing HST surveys. Specifically, we used the
resolved image statistics from the Groth Strip, a deep HST survey
(Ebbels 1998, Rhodes et al 1999). This HST survey sampled at 0.1
arcsec effectively gives us the unsmeared (i.e. before convolution
with ground-level seeing) ellipticities and diameters of an ensemble
of galaxies. The Groth Strip is a set of 28 contiguous pointings in
$V$ and $I$; it covers an area of approx. 108 arcmin$^2$ in a 3'.5
$\times$ 44'.0 region. The magnitude limit is $I \simeq 26$, and the
strip includes about 10,000 galaxies. We use a SExtractor catalogue
derived from the entire strip by Ebbels (1998), containing magnitude,
diameter, and ellipticity for each object.

We model the multi-dimensional probability distribution of galaxy
properties (ellipticity - magnitude - diameter) sampled by this
catalogue, and draw from it a catalogue of galaxies statistically
identical to the Groth strip distribution. We normalise to the median
number density acquired in our observed WHT fields, and spatially
distribute the galaxies with a uniform probability across the field of
view. Star counts with magnitude are modelled from the WHT data
itself, since the Groth strip does not contain enough stars to create
a good model.

We then shear the galaxies in our prospective simulation catalogue
by calculating the change in the object ellipticity due to
lensing. Here we use the relation (Rhodes et al 1999):

\begin{equation}
e'_i = e_i + 2(\delta_{ij} - e_i e_j)\gamma_j
\end{equation}

For the purposes of this paper, we ran 3 sets of simulations: the
first was a null test, with zero rms shear entered for 20 fields; the
second included a 1.5\% rms shear for 30 fields; and the third a 5\%
rms shear for 20 fields. This will allow us to check the KSB method in
the weak shear-measuring regime. The imposed shear is uniform over a
given field; this simplification should not affect our results, since
we are primarily interested only in the mean shear measured on the
field. We select uniform shears for each field from a Gaussian
probability distribution with standard deviation equal to the rms
shear we wish to study.

Stellar ellipticities (i.e. tracking errors) are similarly chosen as
uniform over a given field, taken from a Gaussian probability
distribution with $\sigma$=0.08 (this is conservatively chosen to be
slightly worse than the rms stellar ellipticity of the stars in our
data, with $\sigma$=0.07). 

\vspace{.4 cm}

We create the catalogue using the IRAF {\tt artdata} package. This
takes the star and (sheared) galaxy catalogues, and plots the
objects at the specified positions with specified ellipticity,
magnitude, diameter and morphology (only exponential discs and de
Vaucouleurs profiles are supported; we input the appropriate
proportion of spirals and ellipticals from HST morphological
counts, and model irregulars as de Vaucouleurs profiles).

\vspace{.4 cm}

\begin{figure}
\psfig{figure=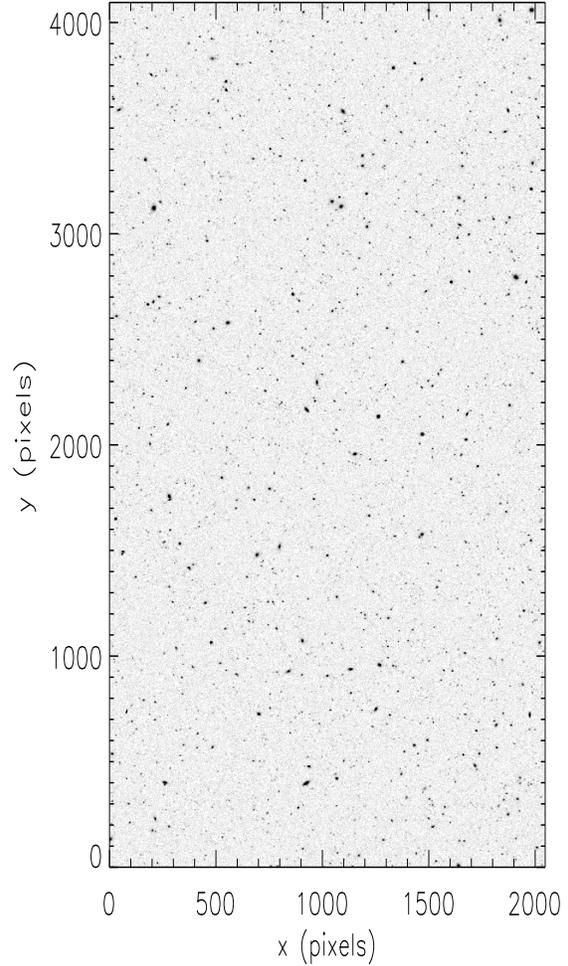,width=80mm,height=140mm} 
\caption{An example simulated image (see text for details).}
\label{fig:simimage}
\end{figure}

We use the package to recreate several WHT-specific details: the
magnitude zero point is chosen to match the telescope throughput, the
stars and galaxies are convolved with the chosen elliptical PSF
(seeing chosen to be 0.8''), the image is appropriately pixelised
(0.237'' per pixel), Poisson and read noise (3.9 electrons) are added,
the appropriate gain (1.45 electrons / ADU) is included, and an
appropriate sky background (10.7ADU per sec) is imposed. The PSF
profile chosen is the Moffat profile, $I(r) = (1 + (2^{1/\beta}-1)
(r/r_{\rm scale})^2)^{-\beta}$, where $\beta$ = 2.5 and $r_{\rm
scale}$ is the seeing radius; $r$ is the distance from the centroid,
transformed so that the profile is elliptical. This profile has wings
which fall off more slowly than for a gaussian profile, and provides a
good description of our seeing-dominated PSF. An example 16' $\times$
8' simulated field is shown in figure \ref{fig:simimage}.

\vspace{.4 cm}

\begin{figure}
\psfig{figure=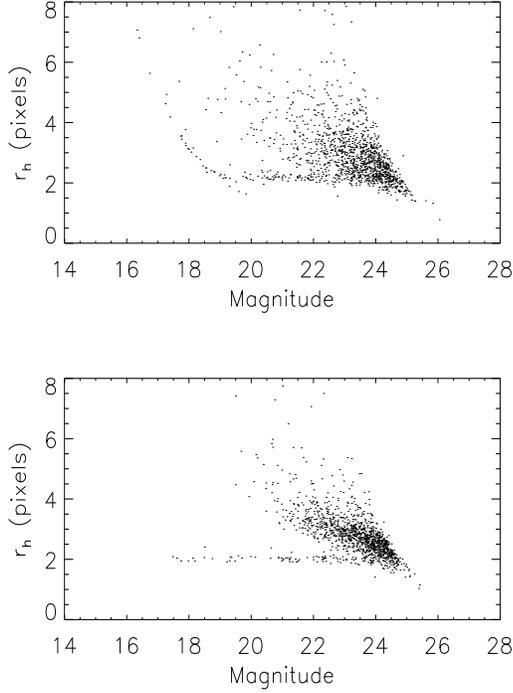,width=80mm,height=100mm}
\caption{The distribution of image magnitudes and half-light radii,
$r_h$, for the data (top) and the simulation (bottom). These
distributions are used for star/galaxy separation.}
\label{fig:magrg}
\end{figure}

Once the simulated catalogues have been realised as images, we run
these through our shear-measurement algorithm, exactly as we did for
the data (see \S\ref{object} and \S\ref{shear_method}). As for the
data, we use 8'$\times$8' cells for shear detection and
measurement. Figure \ref{fig:magrg} demonstrates some of the
similarity between the observed and simulated fields' {\tt imcat}
catalogues.

\vspace{.4 cm}

\begin{figure}
\psfig{figure=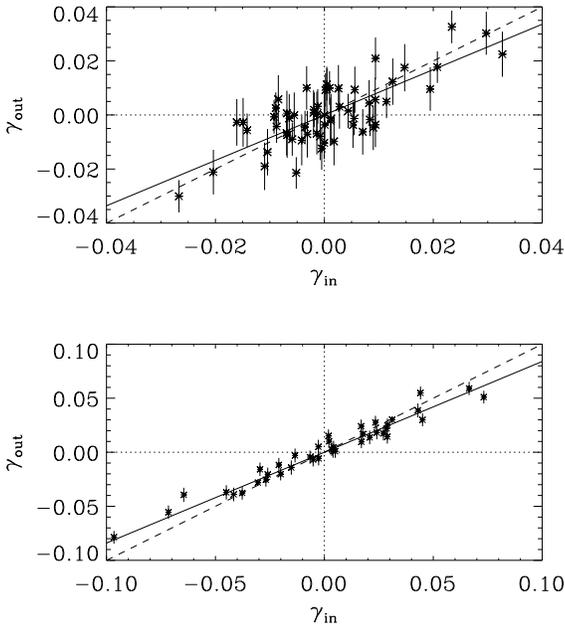,width=80mm}
\caption{$\gamma^{{\rm in}}_{i}$ compared with $\gamma^{{\rm out}}_{i}$ 
for simulated data sheared (top) by 1.5\% rms shear; (bottom) by 5\% rms shear.
The dashed line shows the $\gamma^{{\rm in}}_{i}  = \gamma^{{\rm out}}_{i}$
relation; the solid line shows the best fit, $\gamma^{{\rm in}}_{i}  = 0.84
\gamma^{{\rm out}}_{i}$. }
\label{fig:gingout}
\end{figure}

The next check is a comparison of the input shear for our cells
against the output shear derived by the KSB method; our results for
the 1.5\% and 5\% rms simulations are shown in
figure~\ref{fig:gingout}. The figure shows that the output shear is
clearly linearly related to the input shear, with a slope close to
1. As a quantitative test, we apply a linear regression fit to both
components of the shear combined. For the 5\% rms simulations we
obtain $\gamma^{{\rm out}}_{i} = 0.0007 + 0.84 \gamma^{{\rm in}}_{i}$,
with standard errors on the coefficients of 0.001 and 0.04,
respectively. For the 1.5\% simulations we similarly obtain
$\gamma^{{\rm out}}_{i} = 0.0001 + 0.79 \gamma^{{\rm in}}_{i}$ with
respective standard errors of 0.001 and 0.091. We see that the {\tt
imcat} measure of shear is symmetrical about zero, but appears to
measure the shear as somewhat too small; we therefore adjust our shear
measures by dividing by $0.84\pm0.04$, and account for the uncertainty
in our results analysis.

\vspace{.4 cm}

For low signal-to-noise galaxies, the simulations also display the
anticorrelation between the ellipticities of the galaxies and of the
stars (see \S\ref{anticorr}). For a $\nu>5$ cut, the amplitude of this
anticorrelation is consistent with that found in the data.  As in the
data, the anticorrelation is no longer significant for a $\nu>15$
cut. This confirms both the use of the simulations to test for
systematic effects, and the validity of our signal-to-noise cut. 

\section{Estimator for the Cosmic Shear}
\label{estimator}
\subsection{Shear Variance}
The amplitude of the cosmic shear can be measured by considering the
shear variance in excess of noise and systematic effects.  In our
experiment, we consider $N_{f}$ $8'\times 16'$ fields subdivided into
$N_{c}=2 N_{f}$ $8'\times8'$ cells (see Figure~\ref{fig:cirsi2}).  Let
$\gamma^{fc}_{i}$ be the shear measured in cell $c$ of field
$f$. Here, $c=t$ or $b$ for the top or bottom cells in each field,
respectively. This shear is a sum of the contributions from lensing,
noise and residual systematic effects, and can thus be written as
\begin{equation}
\gamma^{fc}_{i}=\gamma_{i}^{{\rm lens},fc}+\gamma_{i}^{{\rm noise},fc}
       + \gamma_{i}^{{\rm sys},fc}
\end{equation}
We wish to measure the lensing shear variance $\sigma_{{\rm
lens}}^{2} = \langle |\gamma^{{\rm lens},fc}|^{2} \rangle$ in
excess of the noise variance $\sigma_{{\rm noise},fc}^{2} =
\langle |\gamma^{{\rm noise},fc}|^{2} \rangle$ and systematic
variance $\sigma_{{\rm sys,fc}}^{2} = \langle |\gamma^{{\rm
sys},fc}|^{2} \rangle$. An estimator for the lensing variance is
given by
\begin{equation}
\label{eq:sigma_lens_hat}
\widehat{\sigma_{{\rm lens}}^{2}} \equiv \sigma_{{\rm tot}}^{2}
  - \sigma_{{\rm noise}}^{2} - \sigma_{{\rm sys}}^{2},
\end{equation}
where the observed total variance is
\begin{equation}
\label{eq:sigma_tot}
\sigma_{{\rm tot}}^{2} \equiv \frac{1}{N_{c}} \sum_{f,c} |\gamma^{fc}|^{2},
\end{equation}
and the mean noise and systematic variances are defined by
\begin{equation}
\label{eq:sigma_noise}
\sigma_{{\rm noise}}^{2} \equiv \frac{1}{N_{c}} \sum_{fc}
  \sigma_{{\rm noise},fc}^{2}, ~~~
\sigma_{{\rm sys}}^{2} \equiv  \frac{1}{N_{c}} \sum_{fc}
  \sigma_{{\rm sys},fc}^{2}.
\end{equation}
It is easy to check that this estimator is unbiased, i.e. that
\begin{equation}
\langle \widehat{\sigma_{{\rm lens}}^{2}} \rangle = \sigma_{{\rm
lens}}^{2},
\end{equation}
where the brackets denote an ensemble average.

We can also compute the variance of $\widehat{\sigma_{{\rm
lens}}^{2}}$ if we assume that the variables follow a gaussian
distribution.  This is a good approximation for $\gamma_{i}^{{\rm
noise},fc}$, since we are considering an average over many
galaxies (about 2000) in a cell. The systematic contribution to
the shear is dominated by the residual anticorrelation discussed
in \S\ref{anticorr} and thus has a distribution which is close to
that of the stellar ellipticities. The stellar ellipticities are
relatively well approximated by a gaussian distribution. In our
case, it is therefore acceptable to take $\gamma_{i}^{{\rm
sys},fc}$ to be gaussian. The lensing shear $\gamma_{i}^{{\rm
lens},fc}$ is however known to be non-gaussian, especially on
scales smaller than 10' below which nonlinear density
perturbations are dominant (e.g.. Jain \& Seljak 1997; Gatzanaga
\& Bernardeau 1998). In principle, higher order correlation
functions are required. These are however difficult to compute
analytically on such small scales (Scoccimaro et al. 1999; Hui
1999), and are at the limit of the resolution of current N-body
simulations (Jain et al. 1998; Barber et al. 1999; White et al.
1999).

We can now compute the variance of the estimator and find
\begin{eqnarray}
\label{eq:sigma_sigma_lens}
\sigma^{2}[ \widehat{\sigma_{{\rm lens}}^{2}} ]
  & = & \frac{1}{N_{c}} \left[ \left( \sigma_{lens}^{2}
   + \sigma_{noise}^{2} + \sigma_{sys}^{2} \right)^{2} + \right. \nonumber \\
  & & ~~~ \left. 2 \left( \sigma_{\times {\rm lens} 1}^{4}
  + \sigma_{\times {\rm lens} 2}^{4} \right) \right],
\end{eqnarray}
where $\sigma_{\times {\rm lens} i}^{2}$ are the cross-correlation
variances between top and bottom cells due to lensing (see
Eq.~[\ref{eq:sigma_cross}]). In deriving this expression, we have
assumed that the noise and systematic effects are uncorrelated
from the top to the bottom cell.  We have also used the following
approximation
\begin{eqnarray}
\sigma_{{\rm noise}}^{2} & \simeq &
\frac{1}{N_{f}} \sum_{f}
  \sigma_{{\rm noise},ft}^{2} \simeq
\frac{1}{N_{f}} \sum_{c}
  \sigma_{{\rm noise},fb}^{2} \nonumber \\
 & \simeq & \left[ \frac{1}{N_{c}} \sum_{f,c}
  \sigma_{{\rm noise},fc}^{4} \right]^{\frac{1}{2}}
\end{eqnarray}
This is valid given the cells were observed in very similar
conditions, and thus the spread in the $\sigma_{{\rm
noise},fc}^{2}$ is small.

Terms with a `lens' subscript in
equation~(\ref{eq:sigma_sigma_lens}) correspond to cosmic
variance, while the other two terms correspond to the uncertainty
produced by noise and systematic effects. If we are initially
interested in a {\it detection} of cosmic shear, it suffices to
test only the null hypothesis corresponding to the absence of
lensing, i.e. to $\sigma_{{\rm lens}}=\sigma_{\times {\rm lens}
i}=0$. The estimator variance relevant for a detection is
\begin{equation}
\label{eq:sigma_sigma_lens_det}
\sigma^{2}[ \widehat{\sigma_{{\rm lens}}^{2}} ]
  \simeq \frac{1}{N_{c}} \left[
    \sigma_{{\rm noise}}^{2} +
    \sigma_{{\rm sys}}^{2} \right]^{2} ~~~{\rm (detection)}.
\end{equation}

\subsection{Shear Cross-Correlation}

An important aspect of our experiment is our ability to test the
the cross-correlation between the shear measured in 2 adjacent
$8'\times 8'$ cells (see Figure~\ref{fig:cirsi2}). As before, let
$\gamma^{ft}_{i}$ and $\gamma^{fb}_{i}$ be the average shear in
the top and bottom portion of the $8'\times 16'$ field $f$,
respectively. The shear cross-correlation variance (see
Eq.~[\ref{eq:sigma_cross}]) is defined by
\begin{equation}
\sigma_{\times {\rm lens}}^{2} \equiv \langle \gamma^{ft}_{i}
\gamma^{fb}_{i} \rangle,
\end{equation}
where the summation convention is used. As before, we assume that
the noise and systematic effects are uncorrelated across the two
cells. An estimator for this quantity is then
\begin{equation}
\label{eq:sigma_cross_hat}
\widehat{\sigma_{\times {\rm lens}}^{2}} \equiv
  \frac{1}{N_{f}} \sum_{f} \gamma_{i}^{ft} \gamma_{i}^{fb}.
\end{equation}
It is again easy to check that it is unbiased, i.e. that
\begin{equation}
\langle \widehat{\sigma_{\times {\rm lens}}^{2}} \rangle
  = \sigma_{\times {\rm lens}}^{2}
\end{equation}
Assuming as before that the fields are gaussian, we can compute the
variance of this estimator and find
\begin{eqnarray}
\label{eq:sigma_sigma_cross}
\sigma^{2}[ \widehat{\sigma_{\times {\rm lens}}^{2}} ]
  & = & \frac{1}{2N_{f}} \left[ \left( \sigma_{lens}^{2}
   + \sigma_{noise}^{2} + \sigma_{sys}^{2} \right)^{2} + \right. \nonumber \\
  & & ~~~ \left. 2 \left( \sigma_{\times {\rm lens} 1}^{4}
  + \sigma_{\times {\rm lens} 2}^{4} \right) \right],
\end{eqnarray}
which equals $\sigma^{2} [ \widehat{\sigma_{{\rm lens}}^{2}} ]$.
For a detection we must rule out the null hypothesis
($\sigma_{\times {\rm lens} 1}^{2}=\sigma_{\times {\rm lens}
2}^{2} = \sigma_{{\rm lens}}^{2}=0$). The relevant estimator
variance for this purpose is then
\begin{equation}
\label{eq:sigma_sigma_cross_det}
\sigma^{2}[ \widehat{\sigma_{\times {\rm lens}}^{2}} ]  =
 \frac{1}{2 N_{f}} \left[ \sigma_{noise}^{2} + \sigma_{sys}^{2}
  \right]^2 ~~~{\rm (detection)}.
\end{equation}

\section[]{Results}
\label{results}

We now present and interpret our results, first using the simulations,
and then examining the WHT data. The following description is
summarised in Table \ref{tab:results} for convenience.

\subsection{Simulated Fields}

\begin{figure}
\psfig{figure=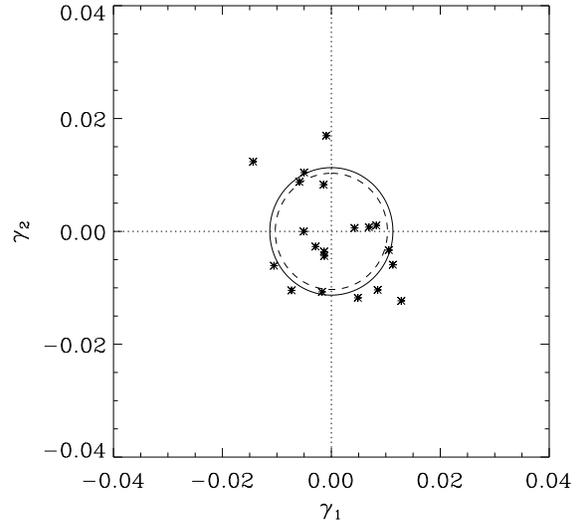,width=80mm}
\caption{Mean $\gamma_1$ and $\gamma_2$ for 20 simulated null
cells. The dashed circle shows the noise rms, the solid circle
shows the total rms. The difference is consistent with zero
signal.}
\label{fig:shearnull}
\end{figure}

We begin with the null simulations, which consists of 20 8'$\times$8'
disjoint cells. The distribution of the shear for each simulated cell
is shown on figure~\ref{fig:shearnull}.

For the null simulation, the rms noise (Eq.~[\ref{eq:sigma_noise}]) is
$\sigma_{{\rm noise}} \simeq 0.0103$, while the observed total
rms is $\sigma_{{\rm total}} \simeq 0.0113$
(Eq.~[\ref{eq:sigma_tot}]) . The noise
and total rms are indicated as a solid and dashed line in
figure~\ref{fig:shearnull}, respectively. Clearly, in the absence of a lensing
signal, $\sigma_{{\rm sys}}^2 = \sigma_{{\rm total}}^2 - \sigma_{{\rm
noise}}^2$, which gives $\sigma_{{\rm sys}} \simeq 0.0047$.

We also require the error for $\sigma_{{\rm sys}}$. In a fashion
similar to that of equation(\ref{eq:sigma_sigma_lens_det}), we find
that
\begin{equation}
\sigma^{2}[ \widehat{\sigma_{{\rm sys}}^{2}} ]
  \simeq \frac{1}{N_{c}} \left[
    \sigma_{{\rm noise}}^{2} +
    \sigma_{{\rm sys}}^{2} \right]^{2}
\end{equation}
giving $\sigma[ \widehat{\sigma_{{\rm sys}}^{2}} ] \simeq
(0.0053)^{2}$, so that $\sigma^2_{{\rm sys}}= (0.0047)^{2} \pm
(0.0053)^{2}$.  Note that this is consistent with zero; i.e. even if
we supposed that there were no systematics, the excess shear signal
that we would attribute to real lensing would be consistent with zero.

We can check this result against our next simulation, which now
includes a 1.5\% rms shear in 30 8'$\times$8' cells. We can first
use this simulation to derive an independent constraint on
$\sigma_{{\rm sys}}$.
\begin{equation}
\widehat{\sigma_{{\rm sys}}^{2}} = \sigma_{{\rm total}}^2 - \sigma_{{\rm noise}}^2 - \sigma_{{\rm lens}}^2
\end{equation}
where we let $\sigma_{{\rm lens}}=0.015$ i.e. the input rms shear.

\begin{figure}
\psfig{figure=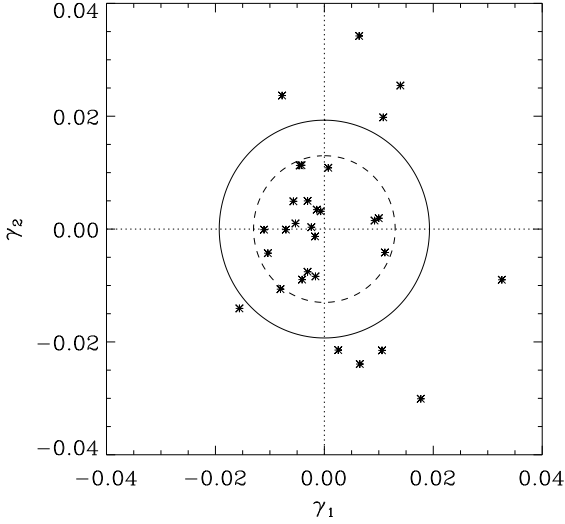,width=80mm}
\caption{Mean $\gamma_1$ and $\gamma_2$ for 30 simulated cells
with rms 1.5\% shear. The dashed circle shows the noise rms, the
solid circle shows the total rms. }
\label{fig:shear15}
\end{figure}

For this simulation, we find $\sigma_{{\rm noise}} \simeq 0.0130$,
$\sigma_{{\rm total}} \simeq 0.0193$ (see
figure~\ref{fig:shear15}). The error for $\sigma_{{\rm sys}}$ this
time is computed as follows
\begin{equation}
\sigma^{2}[ \widehat{\sigma_{{\rm sys}}^{2}} ]
  \simeq \frac{1}{N_{c}} \left[
    \sigma_{{\rm noise}}^{2} +
    \sigma_{{\rm sys}}^{2} + \sigma_{{\rm lens}}^2 \right]^{2}.
\end{equation}
However, since $\sigma_{{\rm total}}^2 - \sigma_{{\rm noise}}^2 <
0.015^2$, we can only find an upper limit for $\sigma_{\rm sys}$ here;
we find that $\sigma[\sigma_{{\rm sys}}^2] = \pm(0.0082)^{2}$,
consistent with the null simulation result. Accordingly, in what
follows, we will use the null simulation estimate for $\sigma^2_{{\rm
sys}}$.

Turning this around, we can use equation~(\ref{eq:sigma_lens_hat}) to
estimate the rms shear in these simulations (ignoring our knowledge of
the input rms shear). We obtain (using the null simulation estimate of
$\sigma_{\rm sys}$) $\sigma_{\rm lens} \simeq 0.013$. The uncertainty in
$\sigma_{\rm lens}^2$ is calculated using
equation~(\ref{eq:sigma_sigma_lens_det}) for detection and
equation~(\ref{eq:sigma_sigma_lens}) for measurement. We obtain
$\sigma[\sigma_{\rm lens}^2]=(0.0060)^{2}$ for detection, and
$\sigma[\sigma_{\rm lens}^2]=(0.0082)^{2}$ for measurement.
Notice that this is the same value as $\sigma[\sigma_{\rm sys}^2]$,
since we can't independently find $\sigma_{\rm sys}$ and $\sigma_{\rm
lens}$ for the simulations.  The measured rms shear is thus fully
consistent with the input rms shear of 1.5\%
(Figure~\ref{fig:shear15}).

An analogous analysis is done for the 5\% rms shear simulations; see
table \ref{tab:results}. Again, we recover the input rms shear within
the uncertainties. Note again that, since $\sigma_{{\rm total}}^2 -
\sigma_{{\rm noise}}^2 < 0.05^2$, only an upper limit can be found for
$\sigma_{\rm sys}$ here. We can conclude that the simulations
clearly show that in the relevant regimes, our method is unbiased.

\subsection{Observed fields}
\begin{figure}
\psfig{figure=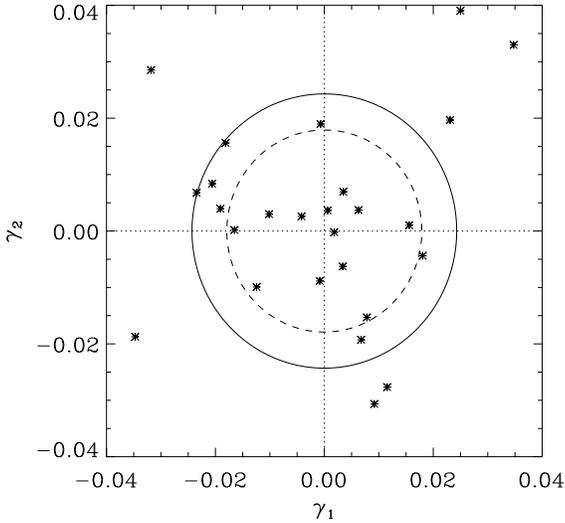,width=80mm}
\caption{Mean $\gamma_1$ and $\gamma_2$ for the observed cells.
The dashed circle shows the noise rms, the solid circle shows the
total rms.}
\label{fig:shear}
\end{figure}

We now consider the observed fields. The distribution of shear for
each of the 26 cells is plotted on figure~\ref{fig:shear}, along
with circles corresponding to $\sigma_{{\rm noise}}$ and
$\sigma_{{\rm tot}}$. The mean shear components are
$\overline{\gamma}_1 = -0.00097 \pm 0.0034$ , $\overline{\gamma}_2 =
0.0021 \pm 0.0034$ , fully consistent
with zero as they should be in the absence of systematic
effects. What is more, we are measuring a total shear
variance in excess of the noise. We now determine whether this
detection is significant.

The value for the rms noise is $\sigma_{{\rm noise}}=0.018$, somewhat
larger than in our simulations. This is due to increased noise from
stellar ellipticity fitting and, in poorer seeing cases, lower number
density. The total rms shear is $\sigma_{{\rm tot}}=0.024$, and using
$\sigma_{{\rm sys}}=0.0047$ from the null simulations, we obtain
$\sigma_{{\rm lens}}=0.0156$ (from Eq.~[\ref{eq:sigma_lens_hat}]).

\begin{table*}
 \centering \begin{minipage}{180mm}
\caption{Shear measurement results for the simulated and WHT fields}
\label{tab:results}
\begin{tabular}{lllll} \hline
& Sim. & Sim. & Sim. & Data\\
& Null & 1.5\% & 5\% & \\
\hline
$N_c$ & 20 & 30 & 20 & 26 \\
$\sigma_{{\rm tot}}^{2}$  & $(0.0113)^2$ & $(0.0193)^2$  &
$(0.0494)^2$ & $(0.0243)^2$\\
$\sigma_{{\rm noise}}^{2}$& $(0.0103)^2$ & $(0.0130)^2$ &
$(0.0102)^2$ & $(0.0179)^2$\\
$\sigma_{{\rm sys}}^{2}$  & $(0.0047)^2\;^a$ & $0\;^c$&
$0\;^d$ & $(0.0047)^2 \; ^e$ \\
$\sigma_{{\rm lens}}^{2}$ & $(0.0047)^2\;^b$ & $(0.0132)^2$ &
$(0.0480)^2$ & $(0.0156)^2$ \\
$\sigma_{\times{\rm lens} }^{2}$ &  &  &  & $(0.0156)^2$ \\
$\sigma[ \sigma_{{\rm sys}}^{2} ]$ & $(0.0053)^2\;^a$ &
$(0.0082)^2$& $(0.0234)^2$ & $(0.0053)^2$ \\
$\sigma[ \sigma_{{\rm lens}}^{2} ]$ (detect) & $(0.0053)^2$
&$(0.0060)^2$ & $(0.0055)^2$ & $(0.0082)^2$ \\
$\sigma[ \sigma_{{\rm lens}}^{2}]$ (measure)& $(0.0053)^2\;^b$
&$(0.0082)^2$ & $(0.0234)^2$ & $(0.0119)^2$\\
\hline
\end{tabular}\\
\footnotesize{$^a$ assumes $\sigma_{{\rm lens}}$=0.}\\
\footnotesize{$^b$ assumes $\sigma_{{\rm sys}}$=0.}\\
\footnotesize{$^c$ assumes $\sigma_{{\rm lens}}$=0.015}\\
\footnotesize{$^d$ assumes $\sigma_{{\rm lens}}$=0.05}\\
\footnotesize{$^e$ Uses the null simulation value, since we can not
obtain an independent estimate}
\end{minipage}
\end{table*}

Using equation~(\ref{eq:sigma_sigma_lens_det}), we find the error in
$\sigma_{{\rm lens}}$ to be $\sigma [ \widehat{\sigma_{{\rm
lens}}^{2}} ] \simeq (0.0082)^{2}$ for the statistical error
only.  If we also include the uncertainty on the systematic (by
adding it in quadrature), we obtain $\sigma[\widehat{\sigma_{{\rm
lens}}^{2}}] \simeq (0.0084)^{2}$. We therefore quote our result as
\begin{eqnarray}
\sigma_{{\rm lens}}^2 &  =  & \sigma_{{\rm lens, measured}}^2  \pm \sigma
[ \widehat{\sigma_{{\rm lens}}^{2}}]_{{\rm statistical}} \pm \sigma
[ \widehat{\sigma_{{\rm sys}}^{2}}] \nonumber \\
& = & (0.0156)^2 \pm (0.0082)^2 \pm (0.0047)^2.
\end{eqnarray}
The significance of our detection of the cosmic shear is therefore
\begin{equation}
(S/N)_{{\rm detect}} = \frac {\sigma_{{\rm lens}}^2}{\sigma
[ \widehat{\sigma_{{\rm lens}}^{2}} ]_{{\rm total}}} \simeq 3.4
\end{equation}

In terms of {\it measuring} the amplitude of the cosmic shear, we use
equation~(\ref{eq:sigma_sigma_lens}) and find $\sigma[\sigma_{\rm
lens}^2] = (0.0119)^{2}$; including the uncertainty on the
systematic we obtain $\sigma[\sigma_{\rm lens}^2] =
(0.0121)^{2}$. We therefore find
\begin{eqnarray}
\sigma_{{\rm lens}}^2 &  =  & \sigma_{{\rm lens, measured}}^2  \pm \sigma
[ \widehat{\sigma_{{\rm lens}}^{2}}]_{{\rm statistical}} \pm \sigma
[ \widehat{\sigma_{{\rm sys}}^{2}}] \nonumber \\
& = & (0.0156)^2 \pm (0.0119)^2 \pm (0.0057)^2,
\end{eqnarray}
where we have included in $\sigma_{{\rm sys}}$ the uncertainty in our
KSB shear calibration (see section \ref{simulation}). Thus we find
that $(S/N)_{{\rm measure}} \simeq 1.7$.

The final measurement we can make is the cross-correlation
covariance using equation~(\ref{eq:sigma_cross_hat}).  We find
that $\sigma_{\times 1} \simeq 0.0115$, $\sigma_{\times 2} \simeq
0.0105$, leading to $\sigma_\times$ = 0.0156. For a detection,
$\sigma[\sigma_\times^2] \simeq (0.0088)^{2}$
(Eq.~[\ref{eq:sigma_sigma_cross_det}]), so that the significance of
the detection is $(S/N)_{{\rm detect}} \simeq 3.2$ for the
cross-correlation. For a measurement, $\sigma[\sigma_\times^2] \simeq
(0.0119)^{2}$ (Eq.~[\ref{eq:sigma_sigma_cross}]), so
$(S/N)_{{\rm measure}} \simeq 1.7$.

\subsection{Cosmological Implications}

A key question we must now address is the redshift distribution of
our background sources. At the median magnitude of the original
catalogue ($R$=25.2$\pm$0.2, Table 2, $\S$3.2), photometric redshift
estimators in various HST and ground-based datasets suggest a mean
redshift $z\simeq$1-1.2 (Fernandez-Soto et al 1999, Poli et al 1999,
Rhodes 1999). More importantly, for the survey sample, which is
effectively limited at a median magnitude of $R\simeq23.4\pm$0.2, we
make use of the recently-completed deep spectroscopic survey of Cohen
et al (2000) which indicates a median redshift at this limit
$z$=0.8$\pm$0.2; the uncertainty here includes the observed
field-to-field variations in this limiting depth as in Table 2.

We can now compare these results with those predicted for the
various cosmological models listed in Table \ref{tab:models}.
First, we compare our value of $\sigma_{\rm lens}^2 = (0.016)^2\pm
(0.012)^{2} $ (with errors which includes cosmic variance). We
find that our result is consistent with the cluster normalised
models $\tau$CDM, $\Lambda$CDM and OCDM at the 0.6-0.9 $\sigma$
level, but that it is inconsistent with the COBE normalised SCDM
model at the 3.0$\sigma$ level. This confirms the fact that the
SCDM model has too much power on small scales when normalised to
COBE.

The cross-correlation variance $\sigma_{\times{\rm
lens}}^{2}=(0.0156)^{2} \pm (0.0119)^{2}$ (again with cosmic
variance included in the uncertainty) does not provide as strong a
constraint. It is consistent with the models, with deviations of
between 0.1$\sigma$ to 1.4$\sigma$. This results from the fact
that $\sigma_{\times{\rm lens}}^{2}$ is expected to have a smaller
amplitude than $\sigma_{{\rm lens}}^{2}$ in all models. It is
nevertheless comforting that, within the context of the models
considered, it is consistent with our measurement of $\sigma_{{\rm
lens}}^{2}$.

We can use our measurement of $\sigma_{{\rm lens}}^{2}$ to constrain
$\sigma_8$, the normalisation of the matter power spectrum on 8
$h^{-1}$ Mpc scales. For the $\Lambda$CDM model with $\Omega_{m}=0.3$,
we find from equation~(\ref{eq:sigg_sig8}) that these two quantities
are related by
\begin{equation}
\sigma_8 = 0.894 z_s^{-0.648} \left(\frac{\sigma_{\rm lens}}{0.01}\right)^{0.8}
\end{equation}
For our value of $\sigma_{\rm lens}$ and setting $z_s = 0.8 \pm 0.2$
(see \S\ref{zref}) and propagating errors, this yields
\begin{equation}
\sigma_8 = 1.47 \pm 0.24 \pm 0.46 = 1.47 \pm 0.51,
\end{equation}
where the first error arises from the uncertainty in $z_s$ and the
second from that of $\sigma_{\rm lens}^2$. This corresponds to a
2.9$\sigma$ measurement of $\sigma_8$. We can compare this with
the cluster abundance determination which yield $\sigma_8 = (0.6
\pm 0.1) \Omega_m^{-0.53} = (1.13 \pm 0.19)
\left(\frac{\Omega_m}{0.3}\right)^{-0.53}$. We see that our result
is consistent with this independent determination. Note that the
uncertainty in $z_{s}$ does not dominate our uncertainty for
$\sigma_{8}$.

\section[]{Conclusions}
\label{conclusion} Using 14 $8'\times16'$ fields observed
homogeneously with the WHT, we have detected a shear signal
arising from weak lensing by large scale structure. Neglecting
cosmic variance (to test the null hypothesis corresponding to the
absence of lensing), we find a shear variance in $8'\times 8'$
cells of $(0.016)^2 \pm (0.008)^2 \pm (0.005)^2$, where the
errors correspond to $1\sigma$ statistical and systematic
uncertainties, respectively. This corresponds to a detection
significant at the $3.4\sigma$ level. Including (gaussian) cosmic
variance, the shear variance is $(0.016)^2\pm(0.012)^2$.  This
is consistent with the value expected for cluster-normalised CDM
models ($\sigma_{\rm lens}=(1.0-1.3)\times10^{-2}$). On the other
hand, the COBE-normalised SCDM model is rejected at the
($3.0\sigma$) level. We have verified our results by measuring the
cross-correlation of the shear in adjacent cells. We find that the
resulting cross-correlation variance for detection is
$(0.016)^2\pm(0.009)^2$, and for measurement is
$(0.016)^2\pm(0.012)^2$, in agreement with that expected in
cluster normalised CDM models. This is consistent with all the
models considered at the $1\sigma$ level.

Our measurement was derived after a careful accounting of the
systematic effects which can produce a spurious shear signal.  We
find that the most serious systematic effect is the PSF
overcorrection for faint objects in the shear measurement method.
We have shown however that, by keeping only sufficiently bright
objects ($S/N>15$), this effect can be made to be smaller
than the statistical uncertainty. Our methods have been tested
using detailed numerical simulations of the shear signal from
appropriately-constructed synthetic sheared images. We find very
good statistical agreement between the simulated and the observed
data. An extensive description of the simulations will be
described in Bacon et al. (2000).

For a given cosmological model, our measurement can be turned into a
measurement of $\sigma_{8}$, the normalisation of the mass power
spectrum on 8 $h^{-1}$ Mpc scales. For a $\Lambda$CDM model with
$\Omega_{m}=0.3$, we get $\sigma_{8}=1.5 \pm 0.2 \pm 0.5$, where the
errors are $1\sigma$ uncertainties resulting from the uncertainty in
the redshift of the background galaxies and from our measurement
error, respectively. This result is consistent with the $\sigma_{8}$
value derived from cluster abundance ($\sigma_{8}=(1.13 \pm 0.19)
\left(\frac{\Omega_m}{0.3}\right)^{-0.53}$, Viana \& Liddle 1996).

The uncertainty in our measurement is clearly dominated by cosmic
variance and statistical errors. This can be improved by
increasing the number of fields $N_{f}$. Since the signal-to-noise
ratio scales as $\sqrt{N_{f}}$, a four fold improvement in $N_{f}$
should yield a $6.8\sigma$ detection and a $3.4\sigma$ measurement
of the rms shear. This, and the presence of other wide field
cameras, offers good prospects for the improvement of the
measurement of $\sigma_{8}$ from cosmic shear. On the other hand
the determination of $\sigma_{8}$ from cluster abundance is
currently measured only at the $~6\sigma$ level and is
fundamentally limited by the finite number of nearby clusters, for
which accurate temperatures can be determined. In addition, it
depends sensitively on the assumption of gaussian initial
conditions. It is therefore likely that cosmic shear measurements
will supplant cluster abundance for the normalisation of the power
spectrum. With an even larger number of fields, one can also
measure the shape of the power spectrum by looking at the
correlation of the shear between and within nearby fields. The
advent of wide-field cameras will make this possible in the near
future.

\section*{Acknowledgements}
We would like to thank Roger Blandford, Chris Benn, Andrew Firth,
Mike Irwin, Andrew Liddle, Peter Schneider, Yannick Mellier, Roberto Maoli and
Jason Rhodes for useful discussions. We are indebted to Nick Kaiser
for providing us with the Imcat software, and to Douglas Clowe for
teaching us to use it. We thank Max Pettini for providing us with one
of the WHT fields. We are also grateful to Ian Smail, the referee, for
useful comments and suggestions. This work was performed within the
European TMR lensing network. AR was supported by a TMR postdoctoral fellowship
from this network, and by a Wolfson College Research
Fellowship.

\bsp

\label{lastpage}

\end{document}